\documentclass[aps,twocolumn,prd,showpacs,showkeys,preprintnumbers,superscriptaddress,nobibnotes,floatfix,longbibliography,notitlepage,nofootinbib]{revtex4-1}

\pdfoutput=1
\usepackage[utf8]{inputenc}
\usepackage{lmodern}
\usepackage{amsmath,amssymb}
\usepackage{graphicx}
\usepackage{hyperref}
\usepackage{url}
\usepackage{color}
\usepackage{enumitem}
\usepackage{subfigure}
\usepackage[dvipsnames]{xcolor}
\usepackage{bm}
\usepackage{epsfig}
\usepackage{amsmath}
\usepackage{amssymb}
\usepackage{slashed}
\usepackage{color}
\usepackage{accents}
\usepackage[dvipsnames]{xcolor}
\usepackage{tikz}
\usepackage{tabularx}

\renewcommand\[{\left[}

\newcommand{\exclude}[1]{}

\usepackage{braket}
\usepackage{url}

\usepackage{mathrsfs}

\newcommand{\tx}{\text}

\usepackage{fontawesome}

\usepackage{comment}

\newcommand{\bp}{\begin{pmatrix}}
\newcommand{\ep}{\end{pmatrix}}
\newcommand{\bb}{\begin{bmatrix}}
\newcommand{\eb}{\end{bmatrix}}

\newcommand{\df}{\text{d}}

\newcommand{\bs}{\boldsymbol}

\newcommand{\al}[1]{\begin{align}#1\end{align}}

\newcommand{\ab}[1]{\left|#1\right|}

\newcommand{\paren}[1]{\left(#1\right)}
\newcommand{\pn}[1]{\left(#1\right)}
\newcommand{\sqbr}[1]{\left[#1\right]}
\newcommand{\fn}[1]{\!\paren{#1}} % function

\newcommand{\wt}{\widetilde}
\newcommand{\nn}{\nonumber\\}

\newcommand{\p}{\partial}

\usepackage{fancybox}

\usepackage{amsthm}
\theoremstyle{definition}

\newcommand{\ov}{\over}
\renewcommand{\ol}{\overline}

\newcommand{\mc}{\mathcal}

\newcommand{\erf}{\text{erf}}

\newcommand{\pr}{\prime}
\newcommand{\eV}{\text{eV}}

\newcommand{\MeV}{\text{MeV}}
\newcommand{\GeV}{\text{GeV}}

\newcommand{\cm}{\text{cm}}

\newcommand{\km}{\text{km}}

\newcommand{\s}{\text{s}}

\usepackage{ulem} % for \sout

\usepackage{fancybox} 

\usepackage{amsthm}
\theoremstyle{definition}
\usepackage{feynmp-auto} 

\usepackage{tikz-feynman}
\tikzfeynmanset{compat=1.1.0}

\renewcommand{\Im}{\text{Im}}

\renewcommand{\min}{\text{min}}
\renewcommand{\max}{\text{max}}

\newcommand{\med}{\text{med}}

\begin{document}

\preprint{KEK-QUP-2026-0009, KEK-TH-2843}

\title{Halo-Independent Quantum Sensor Probes of Low-Velocity Dark Matter}

\author{Muping Chen}
\email{mpchen@post.kek.jp}
\affiliation{International Center for Quantum-field Measurement Systems for Studies of the Universe and Particles (QUP, WPI),
High Energy Accelerator Research Organization (KEK), Oho 1-1, Tsukuba, Ibaraki 305-0801, Japan}

\author{Graciela B. Gelmini}
\email{gelmini@physics.ucla.edu}
\affiliation{Department of Physics and Astronomy, UCLA,\\
475 Portola Plaza, Los Angeles, CA 90095, USA}

\author{Volodymyr Takhistov} \email{vtakhist@post.kek.jp}
\affiliation{International Center for Quantum-field Measurement Systems for Studies of the Universe and Particles (QUP, WPI),
High Energy Accelerator Research Organization (KEK), Oho 1-1, Tsukuba, Ibaraki 305-0801, Japan}
\affiliation{Theory Center, Institute of Particle and Nuclear Studies (IPNS), High Energy Accelerator Research Organization (KEK), Tsukuba 305-0801, Japan
}
\affiliation{Graduate University for Advanced Studies (SOKENDAI), \\
1-1 Oho, Tsukuba, Ibaraki 305-0801, Japan}
\affiliation{Kavli Institute for the Physics and Mathematics of the Universe (WPI), UTIAS, \\The University of Tokyo, Kashiwa, Chiba 277-8583, Japan}
\author{Koichiro Yasuda}
\email{yasuda@physics.ucla.edu}
\affiliation{Department of Physics and Astronomy, UCLA,\\
475 Portola Plaza, Los Angeles, CA 90095, USA}

\begin{abstract}
We present a halo-independent framework for sub-GeV dark matter (DM) direct detection using quantum sensors with sub-eV energy thresholds. Such detectors enable access to low DM velocities and may be sensitive to departures from the Standard Halo Model that are challenging to probe with conventional direct DM detection experiments. The method expresses the DM scattering event rate in terms of a detector  and particle model-dependent response function, and a universal halo function common to all experiments to be determined from data. This allows the local DM velocity distribution to be constrained. 
As representative implementations, we consider TES (Al) and MKID (TiN)-like sensors and show that their differing material responses probe complementary regimes of the DM velocity distribution. Applying the framework to mock data derived from several benchmark local halo models, we demonstrate how the assumed halo function could be reconstructed. This framework demonstrates the potential of quantum sensors as a new avenue for mapping the local DM velocity distribution.
\end{abstract}

\maketitle

%%%%%%%%%%%%%%%%%%%%%%%%%%%%%%%%%%%%%%
\section{Introduction}

The nature of dark matter (DM) is among the central open problems in physics. To date, evidence for DM derives from its gravitational effects on astrophysical and cosmological scales (see  e.g.  Refs.~\cite{Bertone:2004pz, Feng:2010gw} for reviews). Direct detection experiments, which search for energy depositions produced by particle DM interactions in detectors, have placed strong constraints on weak-scale DM candidates~\cite{LZ:2022lsv, XENON:2022ltv, PandaX-4T:2021bab, Akerib:2022ort}.

Increasing attention has therefore turned to new regions of parameter space including light sub-GeV DM, for which conventional nuclear-recoil searches can become kinematically limited~\cite{Essig:2022dfa}. In this regime electronic and collective excitations as well as other low energy processes in condensed matter targets offer promising detection channels~\cite{Essig:2011nj, Hochberg:2015pha, Hochberg:2015fth, Knapen:2017xzo, Griffin:2021znd}. Accessing these small energy depositions requires detectors with eV-scale or sub-eV energy thresholds, low noise and in many cases single quantum sensitivity. These capabilities are increasingly enabled by quantum sensing technologies.

Recent advances in quantum sensing have opened a new experimental window for low threshold DM searches. Representative platforms include transition edge sensors (TESs), an established class of precision calorimeters~\cite{2005cpd..book...63I} that are well suited for light DM detection~\cite{Hochberg:2015fth,Schwemmbauer:2025evp}. Dedicated analysis of high resolution optical TES arrays~\cite{Hattori:2022mze} incorporating realistic detector noise and material response modeling demonstrated their potential for DM searches with thresholds down to $\mathcal{O}(0.1)$~eV~\cite{Chen:2025cvl}. Other promising  platforms include microwave kinetic inductance detectors (MKIDs)~\cite{DayetalMKID,2008JLTP..151..550G,Gao:2024irf}, among other cryogenic and non-cryogenic detector technologies. These platforms offer complementary combinations of low thresholds, energy sensitivity, scalability  and low dark count rates, making them well suited to probing the small energy depositions expected from light DM interactions. 

The observable particle DM interaction signals, however, depend  not only on fundamental particle parameters  but also on the local DM   distribution and the material-dependent response of the target. Disentangling these ingredients is essential for a robust interpretation of either a potential signal or a null result. A significant challenge is that the local DM phase space distribution is not known from first principles~\cite{Read:2014qva, Bovy:2012tw, OHare:2019qxc}. While the Standard Halo Model (SHM)  with a truncated Maxwellian velocity distribution  provides a useful benchmark~\cite{1986PhRvD..33.3495D, 1996APh.....6...87L}, Galactic structure can include streams, substructures or even DM populations stemming from components such as a dark disk~\cite{Read:2008fh, Fan:2013tia,Fan:2013bea, McCullough:2013jma, Randall:2014kta,
Schutz:2017tfp,Widmark:2021gqx,Putney:2025mch,Gould:1987ir,Neufeld:2018slx,McKeen:2022poo}.
These contributions can be especially relevant for low threshold detectors, which probe small energy deposits and therefore are sensitive to low velocity regions of phase space where non-standard DM distributions can strongly affect   predicted interaction rates.

Halo-independent (HI) analysis methods (see e.g.~\cite{Fox:2010bz,Fox:2010bu,Frandsen:2011gi,Gondolo:2012rs,HerreroGarcia:2012fu,Frandsen:2013cna,DelNobile:2013cta,Bozorgnia:2013hsa,DelNobile:2013cva,DelNobile:2013gba,DelNobile:2014eta,Feldstein:2014gza,Fox:2014kua,Gelmini:2014psa,Cherry:2014wia,DelNobile:2014sja,Scopel:2014kba,Feldstein:2014ufa,Bozorgnia:2014gsa,Blennow:2015oea,DelNobile:2015lxa,Anderson:2015xaa,Blennow:2015gta,Scopel:2015baa,Ferrer:2015bta,Wild:2016myz,Gelmini:2015voa, Gelmini:2016pei,Witte:2017qsy,Gondolo:2017jro,Ibarra:2017mzt,Gelmini:2017aqe,Catena:2018ywo}), enable isolating the astrophysical dependence of direct DM detection rates by expressing them in terms of a halo function $\widetilde{\eta}(v_{\rm min})$ where $v_{\rm min}$ is the minimum DM speed required to produce a given recoil signal. Originally developed for nuclear recoil searches, these techniques have also been extended to DM electron scattering~\cite{Chen:2021qao, Chen:2022xzi,Bernreuther:2023aqe}.
In the present context, the HI framework is particularly useful because detector effects can be captured by a response kernel, while the dependence on the local DM velocity distribution remains encoded in $\widetilde{\eta}(v_{\rm min})$. This preserves a linear relation between possible measured rates and the halo function, enabling direct reconstruction and comparison across low threshold detector technologies.

In this work, we present the HI framework for sub-eV and eV energy depositions from DM scattering with electrons in detectors enabled by quantum sensing. We express the event rate in terms of the halo function and formulate a detector level linear response that incorporates finite energy resolution, threshold effects  and binned data observables. As benchmarks, we analyze TES and MKID type sensors and identify the DM speed regions they probe for sub-GeV  DM masses. In addition, using mock data, we demonstrate projected reconstructions of $\widetilde{\eta}(v_{\rm min})$ for the SHM and benchmark nonstandard components, illustrating how low threshold sensors can provide complementary sensitivity to low velocity DM structures that are difficult to access with conventional recoil searches, if a putative DM signal is observed.

This paper is organized as follows. In Sec.~\ref{sec:DMscattering}, we introduce the DM electron scattering formalism relevant for quantum sensor targets. In Sec.~\ref{sec:detectors}, we describe the TES and MKID benchmark configurations and their material response modeling. In Sec.~\ref{sec:HIframework}, we introduce HI formalism and define the halo function $\widetilde{\eta}(v_{\rm min})$. In Sec.~\ref{sec:responsefunctions}, we develop the detector response formalism including energy resolution and threshold effects. In Sec.~\ref{sec:data-analysis}, we define the benchmark halo scenarios, mock data event rates,  binned observables,  response matrix elements  and show projected reconstructions for the benchmark halo functions. We conclude in Sec.~\ref{sec:conclusion}.

\section{Dark Matter Scattering}
\label{sec:DMscattering}

 The total DM interaction rate per unit detector mass is
\begin{equation}
    R  = {1\ov \rho_T}{\rho_\chi\ov m_\chi}\int\df^3 \bs v f_\chi\fn{\bs v}\Gamma\fn{\bs v},
    \label{R-definition}
\end{equation}
where $\rho_T$ is the target mass density, $\rho_\chi$ is the local DM mass density, $m_\chi$ is the DM particle mass, $f_\chi\fn{\bs v}$ is the local DM velocity distribution in Earth's frame  normalized to 1, 
$\int \df^3\bs v f_\chi(\bs v)=1$, and $\Gamma\fn{\bs v}$ is the scattering rate for an incident DM particle velocity $\bs v$.

In this work, we focus on DM scattering off target Standard Model (SM) electrons. Although we do not restrict ourselves to a specific underlying model, interactions of this form can readily arise in scenarios with a vector mediator of mass $m_V$ and couplings $g_\chi$ and $g_e$ with the DM and electrons respectively, such as with a kinetically mixed dark photon~\cite{Holdom:1985ag,Alexander:2016aln}.
The  scattering rate is
\begin{equation}
\Gamma\fn{\bs v}=
{\pi \ol{\sigma}_e\ov \mu_{\chi e}^2}
\int{\df^3\bs q\ov \pn{2\pi}^3}
\ab{\mc F_\med\fn{q}}^2
S\fn{q,E_e},
\label{Gamma of vecv}
\end{equation} 
where $q=\lvert\bs q\rvert$ denotes momentum transfer, $ \mu_{\chi e}$  is the DM-electron reduced mass, and $  S(q,E_e)$   is the material-dependent dynamic structure factor for electronic excitations. The deposited energy is determined by the non-relativistic scattering kinematics $E_e
\simeq
\bs q\cdot\bs v- q^2/(2m_\chi)$,  as explained in App.~\ref{changeofvar}. 

We define the reference DM-electron interaction cross section, evaluated at the reference momentum transfer $q_0$, as
\begin{equation}
\ol{\sigma}_e
=
{g_e^2 g_\chi^2\mu_{\chi e}^2
\ov
\pi\pn{m_V^2+q_0^2}^2}.
\label{refSigma}
\end{equation}
The corresponding interaction mediator form factor, that takes into account its propagator, is
\begin{equation}
\mc F_\med\fn{q} = {m_V^2+q_0^2\ov m_V^2+q^2}.
\label{DM form factor}
\end{equation}
For DM-electron scattering, we adopt the conventional reference momentum $ q_0=\alpha m_e$ following the convention of Ref.~\cite{Essig:2011nj,Chen:2021qao}, where $ m_e$  is the electron mass and   $\alpha$  is the electromagnetic fine structure constant.
For the heavy mediator limit $\mc F_\med\fn{q} \simeq 1$, while for the light mediator limit $\mc F_\med\fn{q} \simeq (q_0/q)^2$.

Imposing energy conservation for the transferred energy $E$, the differential event rate is given by (see the detailed derivation in App.~\ref{detailedEventRate})
\begin{equation}
    {\df R\ov \df E} =
     {1\ov {4\pi}\rho_T\mu_{\chi e}^2}
    \int_{0}^{\infty} \df q~
    q\ab{\mc F_\med\fn{q}}^2  S\fn{q, E}
    \wt\eta\fn{v_\min\fn{q, E}}.
    \label{dRdE}
\end{equation}
The electronic dynamic structure factor $S(\bs q,E)$ depends on intrinsic properties of the target material and encodes material response to a momentum transfer $\bs q$ and an energy transfer $E$. 

For interactions that couple to the electronic charge density through the longitudinal electromagnetic response of the medium, the same structure factor can be expressed in terms of the longitudinal dielectric function $\epsilon_L(E,\bs q)$~\cite{Knapen:2021bwg,Chen:2022xzi}
\begin{equation} \label{eq:structure}
S(\bs q,E) = 
\frac{q^2}{2\pi\alpha} 
\frac{1}{1-e^{-\beta E}} 
\Im\left[
-\frac{1}{\epsilon_L(E,\bs q)}
\right],
\end{equation}
where $\beta=(k_{\rm B}T)^{-1}$, $T$ is the target temperature, and $k_{\rm B}$ is the Boltzmann constant. The quantity
$\Im[-1/\epsilon_L(E,\bs q)]$ is the energy loss function.  

When a measured or computed momentum-dependent dielectric response is unavailable, the electronic response of a metallic target may be approximated in a simplified manner by that of an isotropic homogeneous electron gas. In this approximation, we use the zero-temperature Lindhard dielectric function~\cite{Dressel_Gruner_2002,Fetter:2003,Quinn:2018} with a phenomenological damping width $\gamma$, introduced through the replacement $E\rightarrow E+i\gamma$,
\begin{equation}
\epsilon_{\rm Lind}(E,q)=
1+\dfrac{3\omega_p^2}{q^2v_F^2} 
\Xi(E,q),
\end{equation}
where
\begin{equation}
\begin{aligned}
\Xi 
={}&
\dfrac{1}{2}
+\dfrac{k_F}{4q}
\left[
1-
\left(
\dfrac{q}{2k_F}
-Q_L
\right)^2
\right]
\ln\left[
\dfrac{
\dfrac{q}{2k_F}
-Q_L+1
}{
\dfrac{q}{2k_F}
-Q_L-1
}
\right] \\
&+
\dfrac{k_F}{4q}
\left[
1-
\left(
\dfrac{q}{2k_F}
+Q_L
\right)^2
\right]
\ln\left[
\dfrac{
\dfrac{q}{2k_F}
+Q_L+1
}{
\dfrac{q}{2k_F}
+Q_L-1
}
\right]~,
\end{aligned}
\end{equation}
where $Q_L = (E+i\gamma)/(qv_F)$.
Here, $k_F$ and $v_F$ are the Fermi momentum and Fermi velocity, respectively, $\omega_p$ is the plasma frequency. Within the free-electron approximation $v_F=k_F/m_e$, $k_F=\sqrt{2m_eE_F}$,
$n_e=k_F^3/(3\pi^2)$,
$\omega_p=\sqrt{4\pi\alpha n_e/m_e}$,
where $E_F$ is the Fermi energy, $n_e$ is the effective conduction electron number density.

The corresponding dynamic structure factor is obtained by substituting $\epsilon_{\rm Lind}(E,q)$ into Eq.~\eqref{eq:structure}. The finite width Lindhard model provides a transparent benchmark for the intraband response of an effective free electron medium.

\section{Quantum Sensor Configurations}
\label{sec:detectors}

A broad range of quantum sensor technologies are being considered and developed for light sub-GeV dark matter searches, including calorimetric, resonant  and threshold counting systems. In this work, we focus on two representative energy resolving configurations. We consider an Al target read out by TESs and  TiN MKIDs. Other sensor technologies can be analyzed with a similar approach and framework  once their material response, detection efficiency  and measured energy response are specified.

We adopt material response models appropriate to each benchmark target. For Al, we use the tabulated dielectric response as computed with the \texttt{DarkELF} package~\cite{Knapen:2021bwg}, which combines experimental optical data
with ab initio calculations. For TiN, for which a comparably established low temperature, momentum dependent energy loss function is not readily available in a suitable form for our analysis, we model the response with a free electron finite width Lindhard function described in Sec.~\ref{sec:DMscattering}.

\subsection{Transition Edge Sensors}

For the TES benchmark, we follow the detector configuration considered in Ref.~\cite{Chen:2025cvl} and take Al to be the active target material, corresponding to the superconducting absorber and collection volume coupled to the TES readout. The electronic response of Al is obtained from the dielectric tables of the \texttt{DarkELF} package~\cite{Knapen:2021bwg}. These tables incorporate both conduction electron and interband contributions, retaining the material-specific response relevant to the energy and momentum transfers considered here.

The tabulated Al energy loss function we consider corresponds to a weighted superposition of Mermin oscillator contributions~\cite{Mermin:1970zz,VOS20166,VOS2025109657}. Each contribution is characterized by an oscillator energy, a damping width and a spectral weight. These parameters are fitted to experimental optical constants and reflection electron energy loss spectroscopy data within the \texttt{CHAPIDIF} framework~\cite{VOS20166,VOS2025109657}. The resulting construction provides an experimentally constrained response at small momentum transfer together with a model based extension to finite momentum transfer. This captures the structure response over the energy range relevant for our analysis.

The Al energy loss function has a dominant plasmon feature near $E \simeq 15~\eV$ and
lower energy interband structure, including a feature near $E \simeq 1.5~\eV$ at the
upper edge of the deposited energy range that we analyze here. The plasmon frequency lies well above
this range and is not relevant for the  scattering interactions analyzed here.
Within the analyzed window the Al response is therefore set by the structured
electron-hole continuum of the data driven Mermin energy loss function, which gives the Al target
richer momentum structure than a free electron model and produces distinct
features of the TES response.

\subsection{Microwave Kinetic Inductance Detectors}

For the MKID benchmark, we consider TiN to be the active target material corresponding to both the superconducting MKID film and the absorber contributions. In contrast to Al, a suitable low temperature, momentum dependent dielectric response for TiN is not readily available in a form that can be directly incorporated into our calculation. We therefore model TiN as an effective free electron medium using the finite width Lindhard dielectric function. 

A material specific TiN energy loss function analogous to the experimentally informed Al response would require dedicated optical or electron energy loss spectroscopy data together with a fit of the corresponding   parameters, for example within a Mermin oscillator model~\cite{Mermin:1970zz,VOS20166,VOS2025109657}. In the absence of such a data set, the finite width Lindhard model provides a well defined and reproducible benchmark for the intraband electronic response.

The model is specified by the TiN mass density $\rho_{\rm TiN}$, the effective Fermi energy $E_F$, and the electronic damping width $\gamma_{\rm TiN}$. We adopt the material parameters~\cite{10.1063/1.1403677,ma8063128} of $\rho_\tx{TiN} = 5.4~$g~cm$^{-3}$, $E_F = 3.94~\eV$ and 
$\gamma_\tx{TiN}  = 0.1 E_F$ as benchmark values. Here, $\rho_{\rm TiN}$ determines the normalization of the event rate per unit detector mass, while $E_F$ sets the characteristic momentum and velocity scales of the effective conduction electron population. The damping width accounts for the finite lifetime of electronic excitations and regulates the corresponding spectral features. Using these parameters, the TiN energy loss function is obtained as described in Sec.~\ref{sec:DMscattering}.  

This construction captures the intraband electron-hole and plasmon response of the effective TiN conduction electron medium and provides a consistent benchmark for evaluating the MKID sensitivity. Material specific interband and crystal band contributions may introduce additional structure in the energy loss function, particularly at low deposited energies. We leave this detailed analysis for future study. The eventual inclusion of these contributions through a dedicated TiN dielectric response calculation would refine the detailed mapping between the measured spectrum and $v_{\min}$. The comparison with the experimentally informed Al response therefore also illustrates how the electronic structure of different target materials can produce complementary sensitivity to the local DM velocity distribution.

\section{Halo-Independent Analysis}\label{sec:HIframework}

A HI analysis separates the dependence of the direct DM detection scattering rate on the local DM velocity distribution from the particle physics, target and detector-dependent response ingredients. The formalism was initially developed for DM-nucleus scattering~\cite{Fox:2010bz,Fox:2010bu,Frandsen:2011gi,Gondolo:2012rs,HerreroGarcia:2012fu,Frandsen:2013cna,DelNobile:2013cta,Bozorgnia:2013hsa,DelNobile:2013cva,DelNobile:2013gba,DelNobile:2014eta,Feldstein:2014gza,Fox:2014kua,Gelmini:2014psa,Cherry:2014wia,DelNobile:2014sja,Scopel:2014kba,Feldstein:2014ufa,Bozorgnia:2014gsa,Blennow:2015oea,DelNobile:2015lxa,Anderson:2015xaa,Blennow:2015gta,Scopel:2015baa,Ferrer:2015bta,Wild:2016myz,Gelmini:2015voa,Gelmini:2016pei,Witte:2017qsy,Gondolo:2017jro,Ibarra:2017mzt,Gelmini:2017aqe,Catena:2018ywo} and was subsequently extended to DM-electron scattering~\cite{Chen:2021qao,Chen:2022xzi,Bernreuther:2023aqe}.
The aim of HI analysis is to infer this common astrophysical function directly from measured rates, without imposing a particular functional form for the local DM distribution. This approach is complementary to the conventional halo-dependent analysis, in which a specific form of velocity distribution $f_\chi(\mathbf{v})$ is assumed and the resulting event rates are used to derive constraints or preferred regions in the DM cross-section and mass parameter space.

For a non-relativistic DM particle of mass $m_\chi$ depositing energy $E_e$ through momentum transfer $\bs q$, the minimum incoming speed required is obtained
when  $\bs q$ and $\bs v$ are parallel (see App.~\ref{changeofvar}), with
\begin{equation}
  v_\min(q, E_e) = \frac{E_e}{q} + \frac{q}{2m_\chi}.
  \label{eq:vmin-q}
\end{equation}
As a function of $q$, this expression attains a unique minimum at
\begin{equation}
  q_* = \sqrt{2m_\chi E_e}, \qquad
  v_* = v_\min(q_*, E_e) = \sqrt{\frac{2E_e}{m_\chi}}.
  \label{eq:qstar-vstar}
\end{equation}
Consequently, for every $v_\min > v_*$ the condition of Eq.~\eqref{eq:vmin-q} admits two
momentum transfer solutions, $q_-(v_\min, E_e) < q_* < q_+(v_\min, E_e)$, with details given in Eq.~\eqref{q+q-}. Both branches must be retained when changing the
integration variable from $q$ to $v_\min$, as detailed in App.~\ref{changeofvar}.

For a specified DM mass and interaction model  the predicted event rate can be written as a convolution of a halo function  
that contains the dependence on the local DM velocity distribution $f_\chi(\mathbf v)$  and a response function that encodes the particle physics, target material  and detector response ingredients, as detailed below. We define the halo function as  
\begin{align} 
    \wt\eta\fn{v_\min} 
    &=
    {\rho_\chi\ol\sigma_e\ov m_\chi}
    {
        \int\df^3 \bs v 
        {f_\chi\fn{\bs v}\ov v}\Theta\fn{v - v_\min}
    } \notag\\
    &= {\rho_\chi\ol\sigma_e\ov m_\chi}
        {
            \int_{v_\min}^\infty \df v 
            {F\fn{v}\ov v}
        }.\label{halofuncDef}
\end{align}  
When directional information is not retained, as in the present analysis, the astrophysical dependence can be expressed in terms of the speed   $v = |\bs v|$ distribution  $F(v)= v^2\int d\Omega_v f_\chi(\mathbf v)$ 
that is obtained by integrating the velocity distribution over directions and does not require $f_\chi(\mathbf v)$ to be isotropic. Since $F(v)\geq0$, the halo function $\widetilde{\eta}(v_{\min})$ is non-negative and non-increasing, with $\widetilde{\eta}(\infty)=0$.
 
The detector frame DM velocity distribution is time dependent because of the Earth's orbital motion around the Sun and its daily rotation. In this work, we consider only the time averaged event rate and therefore use the time averaged halo function
$  \widetilde{\eta}(v_{\min})$. 
Annual and daily modulation signals can be treated within the same HI framework but are left for future study.

The function $\wt\eta(v_\min)$ that maximizes a given likelihood is
piecewise constant and non-increasing, with at most $d-1$ downward steps, where
$d$ is the number of data entries. This was first proven for a specific
(unbinned) likelihood using the Karush-Kuhn-Tucker
conditions~\cite{Fox:2014kua, Gelmini:2015voa}, and later clarified and
generalized through convex geometry
arguments~\cite{Gondolo:2017jro, Gelmini:2017aqe}. Equivalently, any likelihood
can be maximized by a DM speed distribution of the
form~\cite{Gelmini:2017aqe}
\begin{equation}\label{eq:Fdeltas}
  F(v)=\sum_{h=1}^{d-1} F_h~\delta(v-v_h)~,
\end{equation}
with parameters $F_h$ and $v_h$. This follows because any set of predicted rates
can be reproduced by a distribution of this form, and the likelihood is always
maximized for some set of rates. Note that, while the best-fit rates are unique,
the best-fit distribution need not be.

Two sided pointwise bands in the $(v_\min,\wt\eta)$ plane around the best fit can
be defined at any chosen confidence level~\cite{Gelmini:2015voa,Gelmini:2016pei,Gelmini:2017aqe},
although we do not compute confidence bands here. For extended likelihoods, suited
to unbinned data, the best fit $\wt\eta(v_\min)$ is guaranteed to be unique. For
likelihoods that depend only on binned data, such as Poisson or Gaussian, it may
not be the case~\cite{Gelmini:2017aqe}. In the latter case one defines a degeneracy band
containing all functions that maximize the likelihood. When such a band is
present, Wilks' theorem does not apply and a $\chi^2$ distribution with one degree
of freedom cannot be assumed in the large sample limit~\cite{Gelmini:2017aqe}. Confidence levels can then be determined  by Monte Carlo methods.

A key property of $\wt\eta(v_\min)$ is that  for a fixed DM mass and interaction
model, including the mediator form factor and the reference cross-section
convention, it is common to all target materials and detector technologies.
Experiments operating at different detected energies and with different targets
probe different windows in $v_\min$, and each detector's response functions
determine which portions of $\wt\eta(v_\min)$ their data can constrain. Consequently,
experiments observing the same DM signal must reconstruct mutually compatible
halo functions within their overlapping sensitivity regions, up to statistical and
systematic uncertainties. The HI approach therefore provides a direct means of
comparing the results of several experiments without imposing a parametric model for the local DM
velocity distribution.

\section{Detector Response Functions}\label{sec:responsefunctions}
\begin{figure*}[t]
    \centering
    \includegraphics[width=0.45\linewidth]{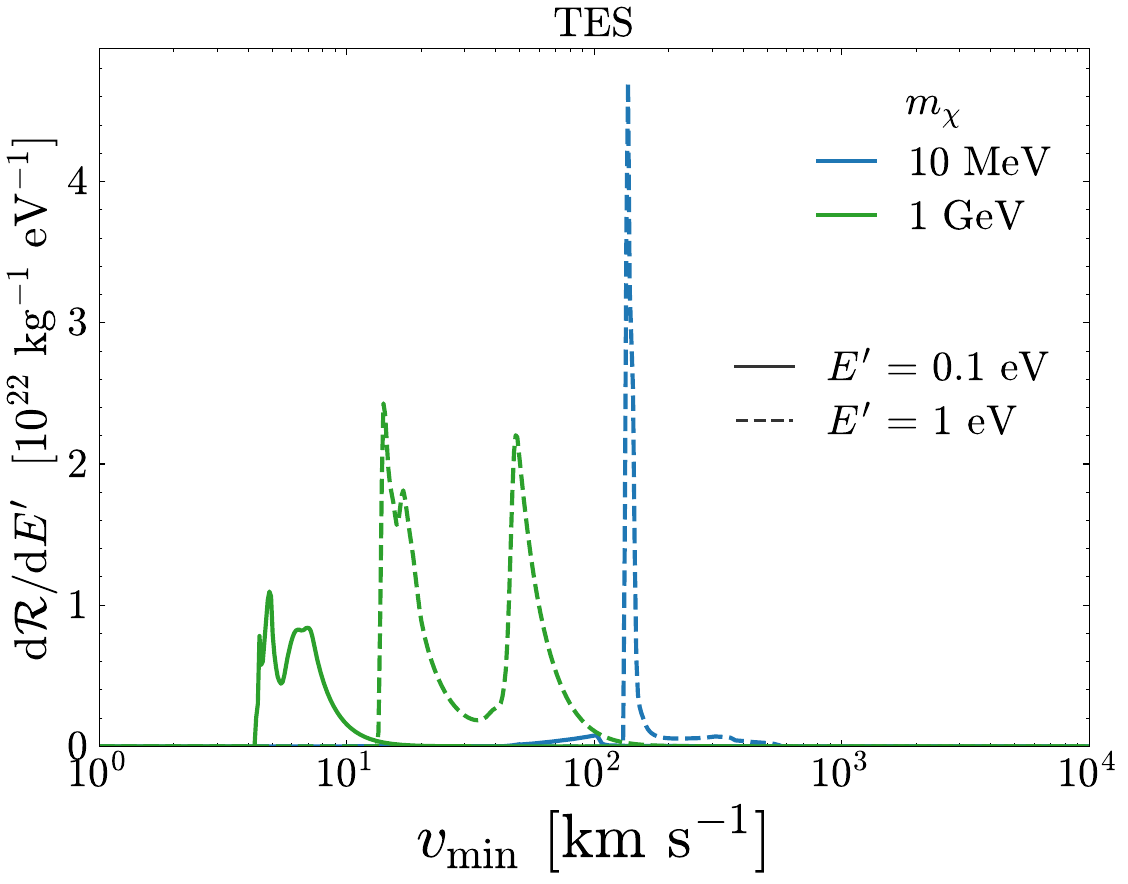}
    \includegraphics[width=0.45\linewidth]{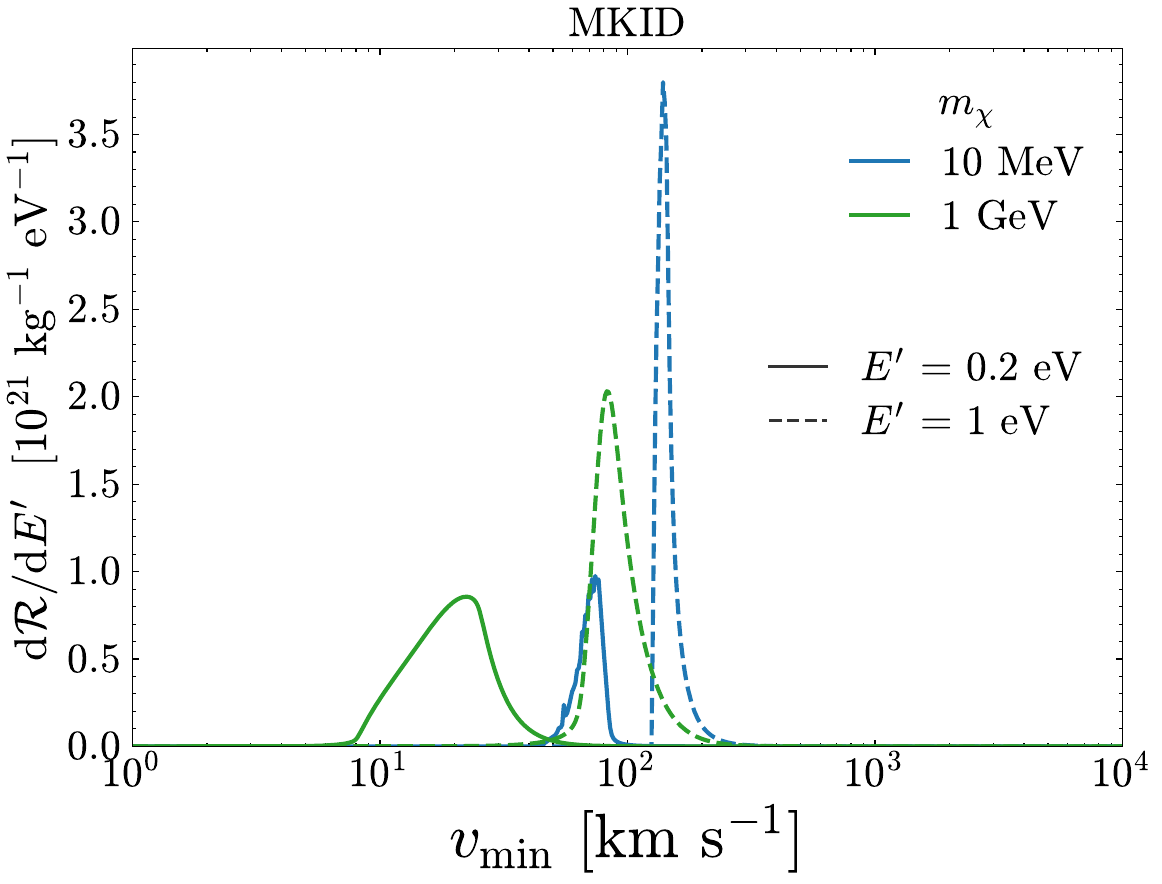}
    \includegraphics[width=0.45\linewidth]{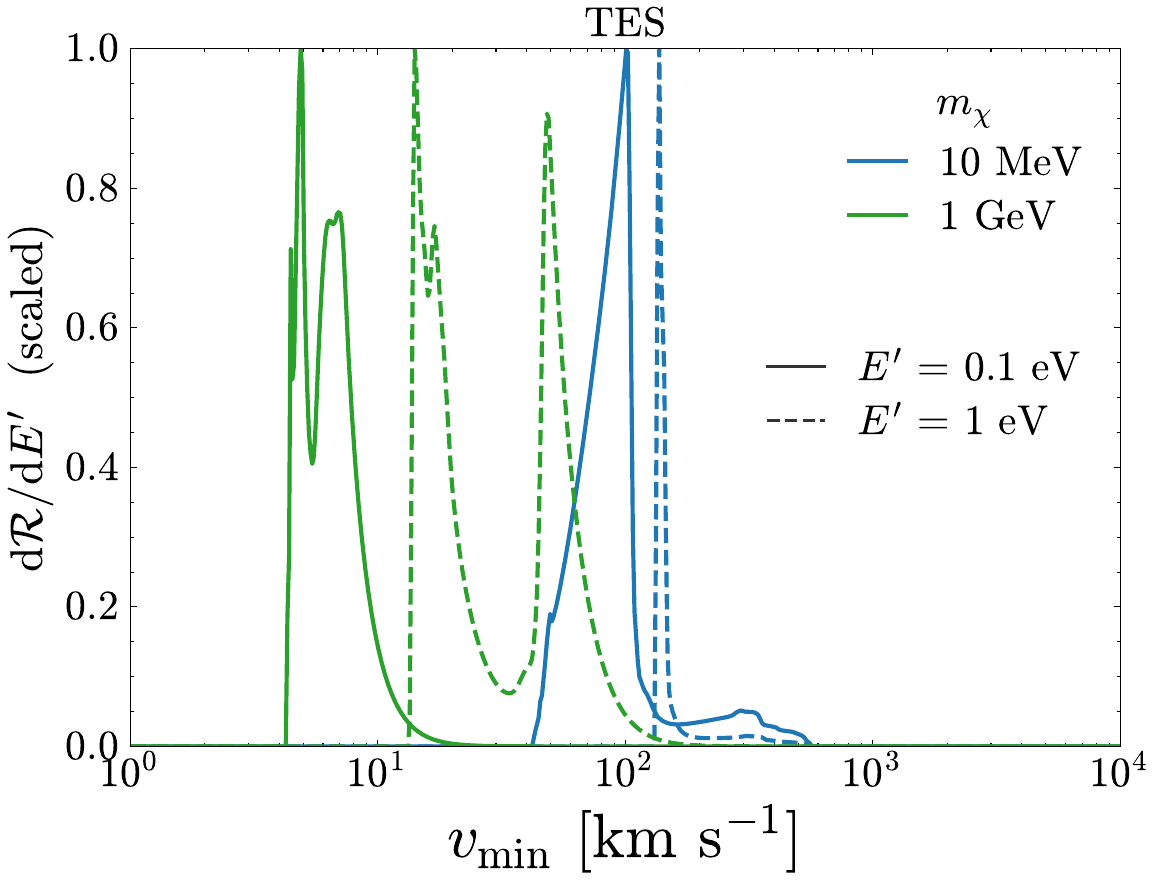}
    \includegraphics[width=0.45\linewidth]{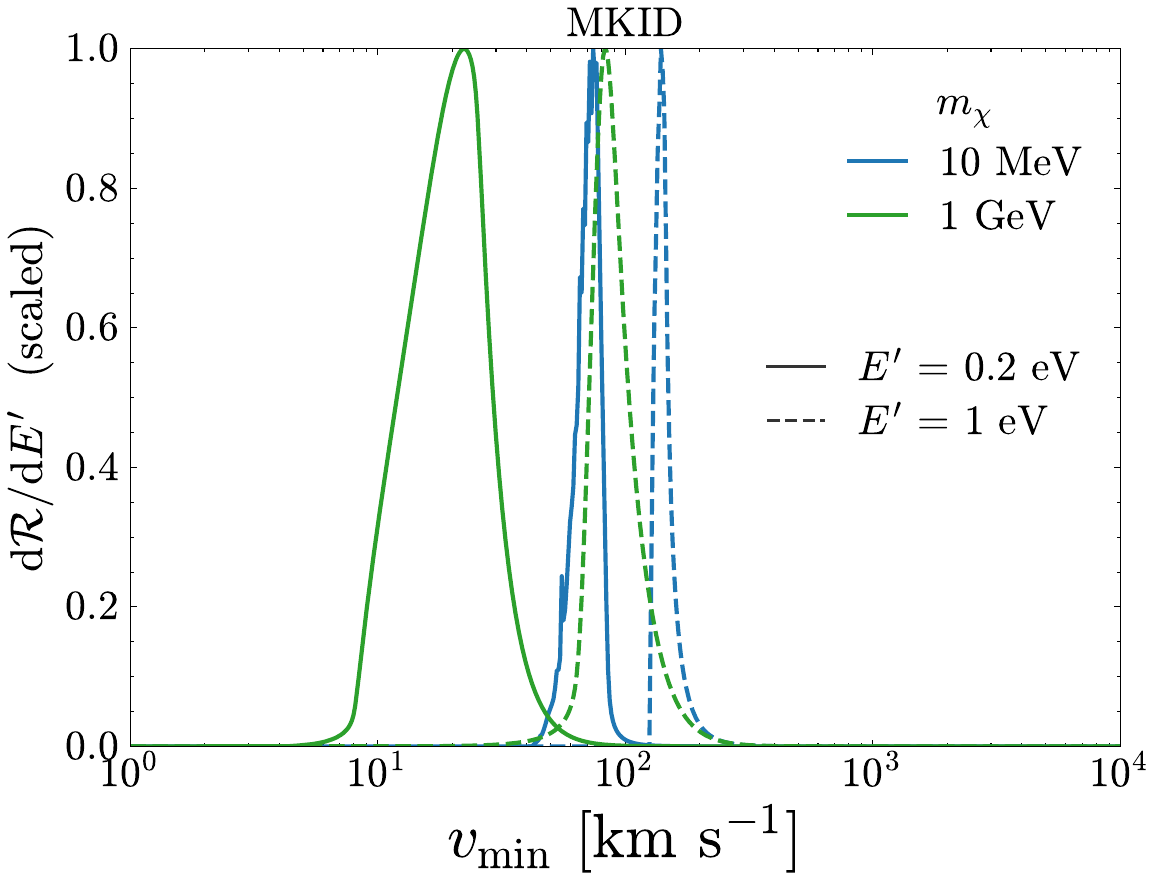}
    \caption{Differential response function $d\mathcal{R}/dE'$ as a function of $v_{\min}$ assuming DM electron scattering with a heavy mediator for the TES (Left) and MKID (Right) detector benchmarks, shown for two representative DM masses  $m_\chi=10$~MeV and $1$~GeV. The actual response functions (Top) are also shown with their maximum normalized to unity (Bottom) to better distinguish the speed ranges where they are considerably non-zero. For TES  the solid and dashed curves correspond to $E'=0.1~\eV$ and $1.0~\eV$, respectively. For MKID, the solid and dashed curves correspond to $E'=0.2~\eV$ and $1.0~\eV$, respectively.}
    \label{fig:differentialcurlyR}
\end{figure*}

\begin{figure*}[t]
    \centering
    \includegraphics[width=0.45\linewidth]{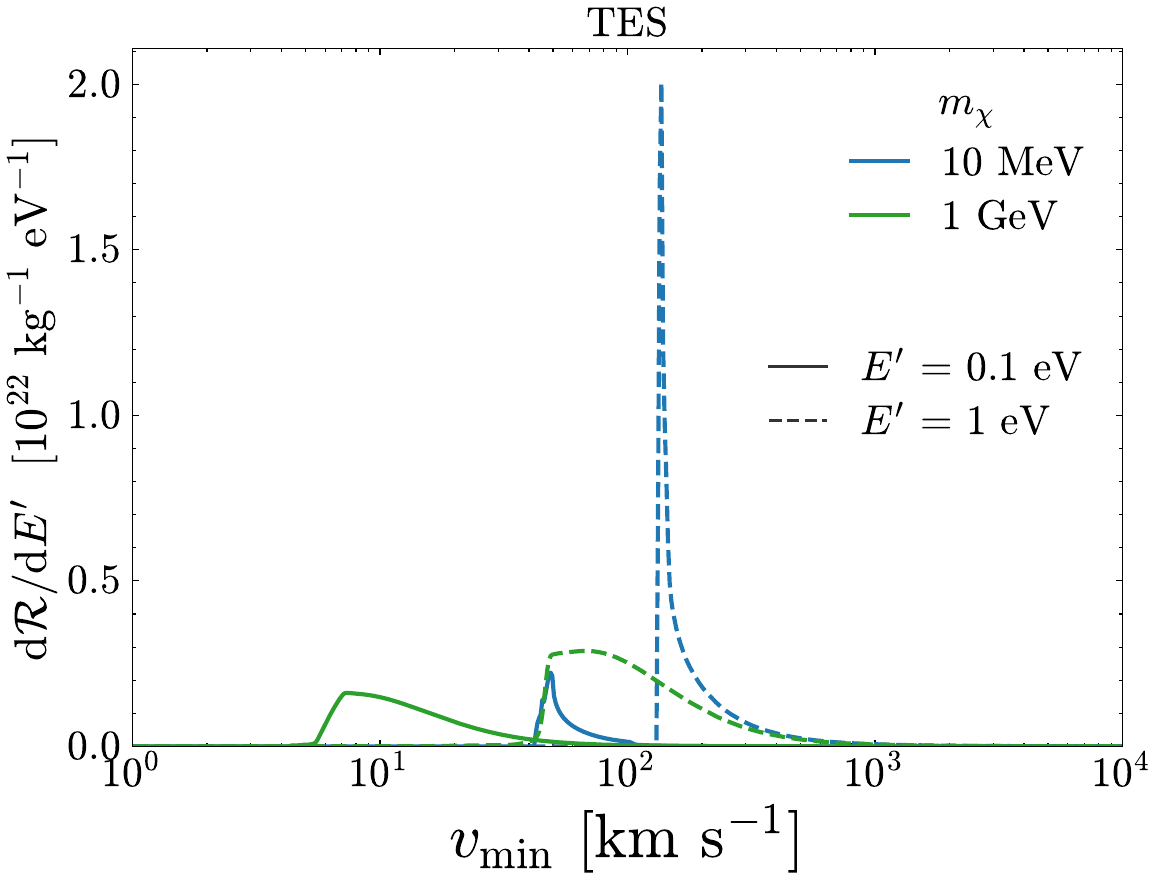}
    \includegraphics[width=0.45\linewidth]{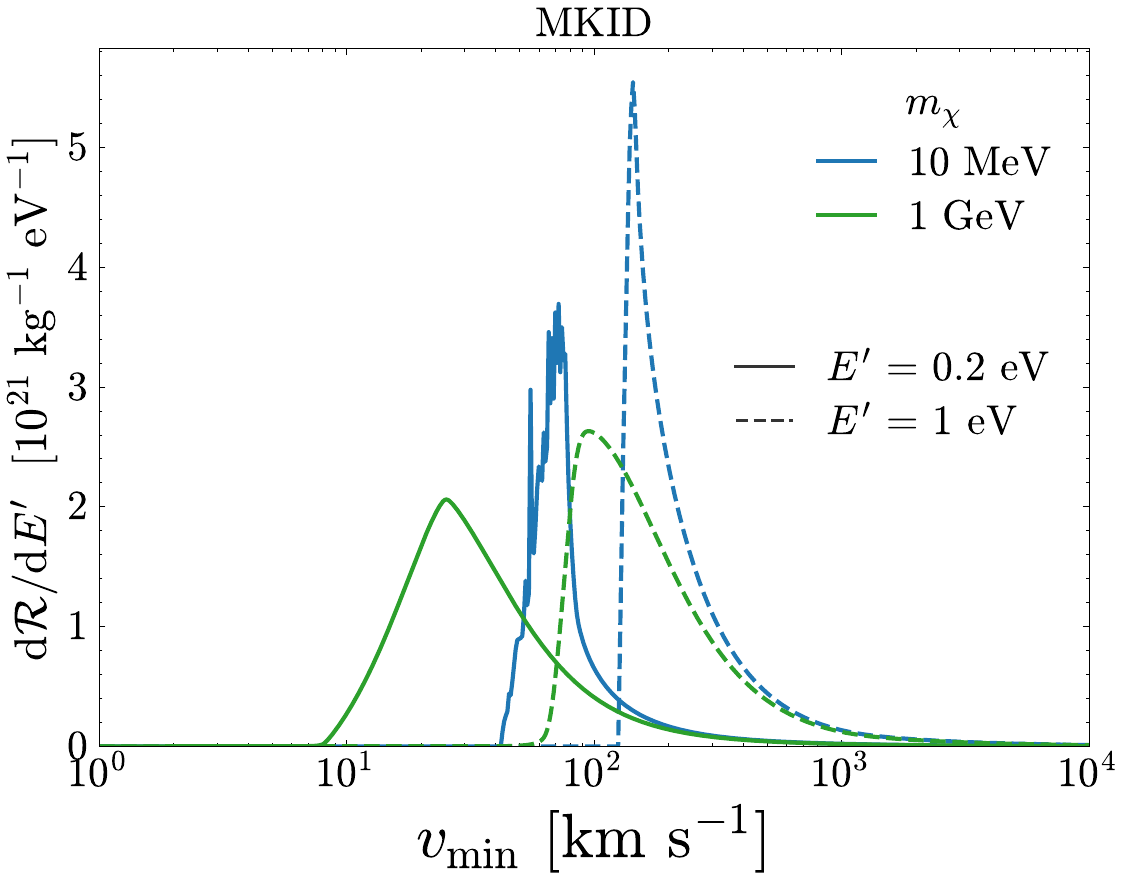}
    \includegraphics[width=0.45\linewidth]{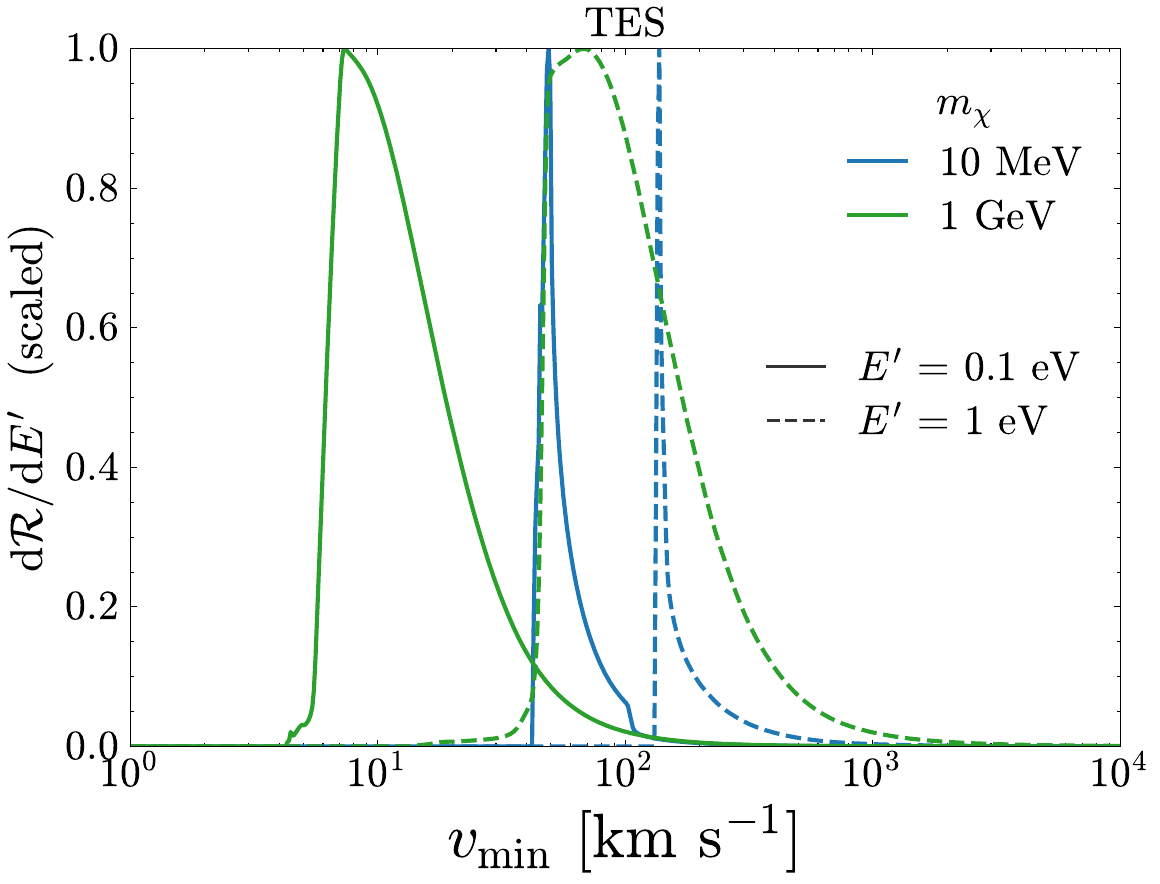}
    \includegraphics[width=0.45\linewidth]{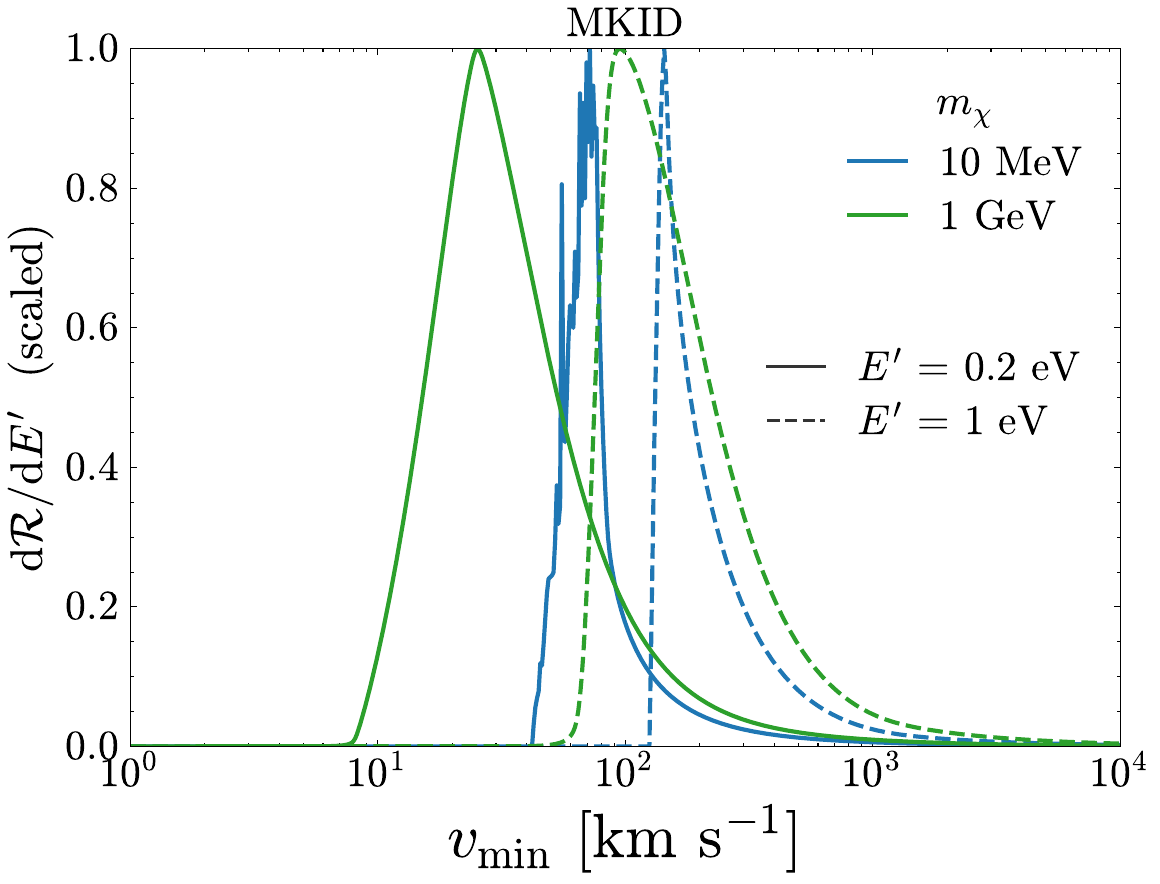}
    \caption{Same as Fig.~\ref{fig:differentialcurlyR}, but assuming DM electron scattering with a light mediator.}
    \label{fig:differentialcurlyRLight}
\end{figure*}

Detectors do not measure the deposited energy $E$ directly  but rather a proxy
$E'$, which may correspond to a number of phonons, photons  or quasiparticles produced in the
detector, which is related to $E$ through an energy resolution function $G(E,E')$ and a
detection efficiency $\varepsilon(E')$. The analysis is therefore carried out in
terms of measurable $E'$, and the response formalism should account for these detector effects. 

We relate the recoil differential rate of Eq.~\eqref{dRdE} to the detected
differential rate  
\begin{equation}
    \frac{\df R}{\df E'}(E')  =
    \varepsilon(E')\int_{0}^{\infty}\df E~  
    G(E,E')  \frac{\df R}{\df E}~.
    \label{eq:observed-diff-rate}
\end{equation}
For simplicity, we assume the detector efficiency can be described as
\al{ \label{eq:threshold}
    \varepsilon(E') &= \theta \left(E' - E'_{\tx{thr}}\right),
}
and a Gaussian energy resolution function of width $\Delta$,
\al{
    G(E,E') &= \frac{1}{\sqrt{2\pi}  \Delta}  
    \exp \left[-\frac{(E-E')^2}{2\Delta^2}\right].
}
For a TES-like and MKID-like experiments we adopt $\Delta_{\tx{TES}} = 0.036~\eV$~\cite{Chen:2025cvl} and  $\Delta_{\tx{MKID}} = 0.13~\eV$~\cite{Gao:2024irf}, respectively. For simplicity, we adopt these values as deliberately broadened Gaussian standard deviation benchmarks  rather than converting the reported full widths at half maximum to Gaussian standard deviations. In practice, we carry out the $E$-integration
in Eq.~\eqref{eq:observed-diff-rate}  numerically, extending
sufficiently far above the center $E'$ of the resolution function $G(E,E')$ for results to converge.

 Our goal is to express the detected rate as a convolution in $v_\min$,
\begin{equation}
    \frac{\df R}{\df E'}(E')
     = \int_{0}^{\infty}\df v_\min  
    \frac{\df \mc R}{\df E'}(v_\min, E')  
    \wt\eta(v_\min),
    \label{dRdEprime1}
\end{equation}
which isolates the astrophysical quantities in $\wt\eta(v_\min)$ and collects all target,
particle physics  and detector dependence into the kernel
$\df\mc R/\df E'$~\cite{Gondolo:2012rs}.
 
We first change the integration variable in Eq.~\eqref{dRdE} from $q$ to
$v_{\min}$ using the two momentum transfer branches $q_\pm(v_{\min},E)$ described in App.~\ref{changeofvar}. For fixed deposited energy
$E$, the map from $q$ to $  v_{\min}$ is two to one for $v_{\min}>v_*(E)$ (see Eq.~\eqref{eq:qstar-vstar}). Splitting the
momentum integral at $q_*=\sqrt{2m_\chi E}$ and including the Jacobian of each
branch gives 
\begin{widetext}
\begin{equation}
\frac{dR}{dE}(E) = 
\frac{1}{4\pi\rho_T\mu_{\chi e}^2}
\int_{v_*(E)}^{\infty} dv_{\min}  \widetilde{\eta}(v_{\min})
\sum_{\lambda=\pm}
\left|\frac{\partial q_\lambda(v_{\min},E)}{\partial v_{\min}}\right|
q_\lambda(v_{\min},E) 
\left|\mathcal F_{\rm med} \left(q_\lambda(v_{\min},E)\right)\right|^2
S \left(q_\lambda(v_{\min},E),E\right).
\label{pureRate}
\end{equation}
\end{widetext} 
The absolute values of the Jacobians account for the opposite orientations of the
two branches with $\partial q_-/\partial v_{\min}<0$, whereas
$\partial q_+/\partial v_{\min}>0$. 

At fixed $v_{\min}$, the deposited energy is
bounded by
\begin{equation}
E\leq E_{\max}(v_{\min})=\tfrac{1}{2}m_\chi v_{\min}^2,
\label{eq:Emax}
\end{equation}
which follows from requiring the momentum-transfer solutions to be real,
$m_\chi^2 v_{\min}^2 - 2m_\chi E \geq 0$. Thus $E_{\max}(v_{\min})$ is the largest
energy a DM particle of minimum speed $v_{\min}$ can deposit considering optimal momentum transfer. Material excitation thresholds are already encoded in the dynamic structure factor $S(q,E)$. 

Inserting the recoil rate $\df R/\df E$ of Eq.~\eqref{pureRate} into
Eq.~\eqref{eq:observed-diff-rate} and exchanging the order of the $E$ and
$v_\min$ integrations, the detected rate becomes
\begin{widetext}
\al{   \label{EventrateDef} 
    {\df R\ov \df E^\pr}&\fn{E^\pr}
    = \varepsilon\fn{E^\pr}\int_{0}^{\infty}\df E  
       G\fn{E, E^\pr}  {\df R\ov \df E} \\
    &= \frac{1}{4\pi\rho_T\mu_{\chi e}^2}
       \int_{0}^{\infty}\df v_\min  \wt\eta\fn{v_\min}  
       \varepsilon\fn{E^\pr}
       \int_{0}^{E_{\max}(v_\min)}\df E  G\fn{E, E^\pr}
       \sum_{\lambda=\pm}
       \ab{\p q_\lambda\ov\p v_\min}  
       q_\lambda\fn{v_\min}  
       \ab{\mc F_\med\fn{q_\lambda\fn{v_\min}}}^2  
       S\fn{q_\lambda\fn{v_\min}, E}, 
}
\end{widetext}
where the deposited energy integration extends up to the kinematic limit
$E_{\max}(v_\min)$ of Eq.~\eqref{eq:Emax}, and the
exchange of integration order maps the lower limit $v_*(E)$ of
Eq.~\eqref{pureRate} onto this upper limit on $E$. Using Eqs.~\eqref{dRdEprime1} and~\eqref{EventrateDef}, one can now identify the differential response function
\al{
    {\df \mc R\ov \df E^\pr}\fn{v_\min, E^\pr}
    &= \sum_{\lambda=\pm}{\df \mc R_\lambda\ov \df E^\pr}\fn{v_\min, E^\pr},
}
with the per branch kernels
\al{
    {\df \mc R_{\pm}\ov \df E^\pr}\fn{v_\min, E^\pr}
    =&~ \frac{1}{4\pi\rho_T\mu_{\chi e}^2}  
       \varepsilon\fn{E^\pr}
       \int_{0}^{E_{\max}(v_\min)}\df E  
       G\fn{E, E^\pr}  
        \nn
    &  \times \ab{\p q_\pm\ov\p v_\min}
       q_\pm\fn{v_\min}  
       \ab{\mc F_\med\fn{q_\pm\fn{v_\min}}}^2  \nn
       & \times S\fn{q_\pm\fn{v_\min}, E}.
}
 
In Figs.~\ref{fig:differentialcurlyR} and~\ref{fig:differentialcurlyRLight}  we show the differential response function $\df\mc R/\df E^\pr$ as a function of $v_\min$ for the TES and MKID configurations for DM electron scattering in the heavy mediator limit and light mediator limit, respectively. The response functions act as window functions in $v_\min$. Measurements at a given observed energy $E^\pr$ constrain the halo function only over the range of $v_\min$ for which the corresponding response are different from zero, shown more clearly in the bottom panels of both figures, with the function normalized to a maximum of approximately 1. The lower edge of this range is not determined solely by the detector threshold. The threshold enters through the efficiency $\epsilon(E^\pr)$ in Eq.~\eqref{eq:threshold}, whereas the $v_\min$ dependence follows from the kinematic minimum
$  v_*$ 
evaluated over the true deposited energies $E$ selected by the resolution function $G(E,E^\pr)$ and weighted by the material response $S(q,E)$. The finite energy resolution therefore smooths and broadens the mapping between $E^\pr$ and $v_\min$.  
 
 For both TES and MKID detector configurations we consider, increasing $E'$ generally shifts the response toward larger $v_\min$, while increasing the DM mass shifts the probed range toward smaller $v_\min$, as can be approximately noted from the scaling $v_*\propto m_\chi^{-1/2}$. The detailed features are target dependent. The Al response used for TES contains several interband and collective excitation features, which produce the multiple peaks visible in Fig.~\ref{fig:differentialcurlyR}, particularly for $m_\chi=1~\GeV$ scenario. The effective free electron TiN response used for MKID is comparatively smoother and selects a different, although partially overlapping, range of $v_\min$. The bottom panels of Fig.~\ref{fig:differentialcurlyR} and Fig.~\ref{fig:differentialcurlyRLight}, in which each response is normalized to its maximum, make these distinct velocity windows and their complementarity more apparent.
 
For the light mediator case since $\mc F_\med(q) \propto q^{-2}$ the  resulting strong weighting toward small momentum transfer suppresses the large $q$ branch relative to the low $q$ branch and thus redistributes the response in $v_\min$. Consequently, both the normalization and the shape of the $v_{\rm min}$ space window depend on the assumed mediator behavior. The HI reconstruction is therefore model independent with respect to the DM velocity distribution, but depends on the chosen DM mass and interaction model.

%%%%%%%%%%%%%%%%%%%%%%%%%%%%%%%%%%%%%%
 \section{Benchmark Halo Reconstruction}\label{sec:data-analysis}

In this section we apply the method developed above to mock data, using the
response functions obtained in Sec.~\ref{sec:responsefunctions}. We generate event rates for
several benchmark local DM velocity distribution models, adopting a realistic local DM density $\rho_\tx{DM} \simeq  0.4~\GeV$~\cm$^{-3}$ (see e.g. Ref.~\cite{2025MNRAS.542.2987S}) together with a deliberately large
reference DM-electron cross section of $\sigma_e = 10^{-30}~\cm^2$.
The HI method is most informative when a sufficient number of signal events is observed to reconstruct an energy spectrum. Hence, to illustrate this, the large cross
section is chosen so that the predicted event counts are sufficiently high to expose the
performance of the reconstruction method itself.
We note that this cross section choice is illustrative. For our analysis we treat the resulting mock event rates
as ``observed'' data and reconstruct the input halo function used to generate
them, assuming a heavy mediator throughout.  

From the analysis we obtain piecewise constant best-fit halo functions as expected. We find that TES-like and MKID-like example
experiments yield complementary results.

\subsection{Dark Matter Velocity Distribution Models}\label{benchmarks}

We analyze the reconstruction of three benchmark local DM velocity distributions. The first is the
Standard Halo Model (SHM), which serves as our baseline. The second is the SHM
with an added co-rotating dark disk (DD). The third is the SHM with an added
Earth-bound (EB) population. The two non-standard benchmarks beyond pure SHM introduce low $v_\min$
structure of the kind that low threshold quantum sensors are particularly well equipped to probe. We
define the three models and their halo functions $\wt\eta(v_\min)$ below and quote
the closed form expressions in App.~\ref{Local-DM}. The corresponding curves and
their reconstructed best fits are shown in Figs.~\ref{SHM}, \ref{DarkDisk}
and~\ref{Bound}.

For the SHM the local distribution is an isotropic Maxwell-Boltzmann in the
Galactic frame, truncated at the Galactic escape speed and boosted to the detector
frame by Earth's motion. In terms of the DM velocity   $\bs v$  in the lab frame velocity,  with the lab
at rest with respect to Earth,
\al{
f_\chi(\bs v) = \frac{1}{N_0}~
e^{-(\bs v + \bs v_e)^2/v_0^2}~
\Theta\!\pn{v_\tx{esc} - |\bs v + \bs v_e|},
\label{SHM_dif}
}
where $N_0$ normalizes the distribution to unity, $v_0$ is the local circular
speed with velocity dispersion $\sigma_0 = v_0/\sqrt{2}$, $\bs v_e$ is Earth's mean
velocity relative to the Galaxy and $v_\tx{esc}$ is the Galactic escape speed at
the solar location. We adopt the conventional values $v_0 = 238~$km~s$^{-1}$, $v_e = 250~$km~s$^{-1}$ and
$v_\tx{esc} = 544~$km~s$^{-1}$~\cite{Baxter:2021pqo}. With these values $\wt\eta_\tx{SHM}(v_\min)$, given in closed form in
App.~\ref{Local-DM}, varies slowly at low $v_{\rm min}$, changes analytic form at $v_\tx{esc}-v_e \simeq
294~$km~s$^{-1}$, and vanishes when $v_\min \geq v_\tx{esc}+v_e \simeq 794~$km~s$^{-1}$ as displayed in Figs.~\ref{SHM}, \ref{DarkDisk} and~\ref{Bound}.

As a first   benchmark beyond SHM we consider an additional contribution from a DD. Such a component
can arise as a relic of the Galaxy's mergers history or in models where a
subdominant DM fraction undergoes dissipative self-interactions and settles into a
rotationally supported structure, which can be approximately co-planar with the baryonic disk and
lagging behind it in rotation~\cite{Read:2008fh, Fan:2013tia, Fan:2013bea,
McCullough:2013jma, Randall:2014kta}. In the Solar neighborhood this results in a small
dispersion and a modest bulk velocity relative to Earth and an excess in
$\wt\eta(v_\min)$ at low $v_\min$, where the SHM is nearly flat. 

We adopt the DD
model of Ref.~\cite{Peter:2011eu}, a truncated Maxwellian form with $v_0^\tx{DD} = 70~$km~s$^{-1}$,  $v_e^\tx{DD} = 100~$km~s$^{-1}$ (which is  Earth's speed with respect to the DD rest frame) and $v_\tx{esc}^\tx{DD} = 694~$km~s$^{-1}$ (which is the escape speed with respect to the DD). Here,
both dispersion and bulk speed are well below their SHM counterparts, so that
$\wt\eta_\tx{DD}(v_\min)$ is concentrated at low $v_\min$. The total halo function
is the density weighted combination
\al{
\wt\eta_\tx{SHM+DD}(v_\min) =
(1 - f_\tx{DD})~\wt\eta_\tx{SHM}(v_\min)
+ f_\tx{DD}~\wt\eta_\tx{DD}(v_\min),
\label{etaSHMpDD}
}
where $f_\tx{DD}$ is the DD fraction of the local density
$\rho_\tx{DM} = 0.4~\GeV$~\cm$^{-3}$ and $\wt\eta_\tx{DD}$ follows from
Eq.~\eqref{halofuncDef} with the DD distribution. Although DD fractions as large
as $\sim25\%$ were previously considered~\cite{Fan:2013bea}, more recent analyses following Gaia survey observations disfavor
values above $\sim5\%$~\cite{Schutz:2017tfp, Widmark:2021gqx, Putney:2025mch}. We display
both $f_\tx{DD} = 5\%$ and $25\%$ cases as illustrative reconstructions.

As a second, more extreme benchmark we add a gravitationally Earth-bound 
population, which can build up as DM particles lose energy in repeated scatterings
in the Earth or Sun and become captured in bound
orbits~\cite{Gould:1987ir, Neufeld:2018slx, McKeen:2022poo}. Earth capture proceeds through DM scattering on SM constituents such
as baryons, thus the capture rate and the resulting bound density are
model-dependent. We do not link this capture to the DM-electron scattering
interaction that we analyze as producing signals in experiments. Instead we treat the bound DM population
phenomenologically remaining agnostic regarding its specific model-dependent particle interaction origin, fixing its density and velocity parameters directly and thus 
isolating its effect on the reconstruction. The assumed bound DM density can be many orders
of magnitude larger than the ambient  local DM density. Its very low lab frame
velocities produce a steep enhancement of $\wt\eta(v_\min)$ as $v_\min \to 0$,
which for sub-GeV DM can be accessed by sub-eV threshold quantum  sensors.

\begin{figure*}[t]
    \centering
    \includegraphics[width=0.32\linewidth]{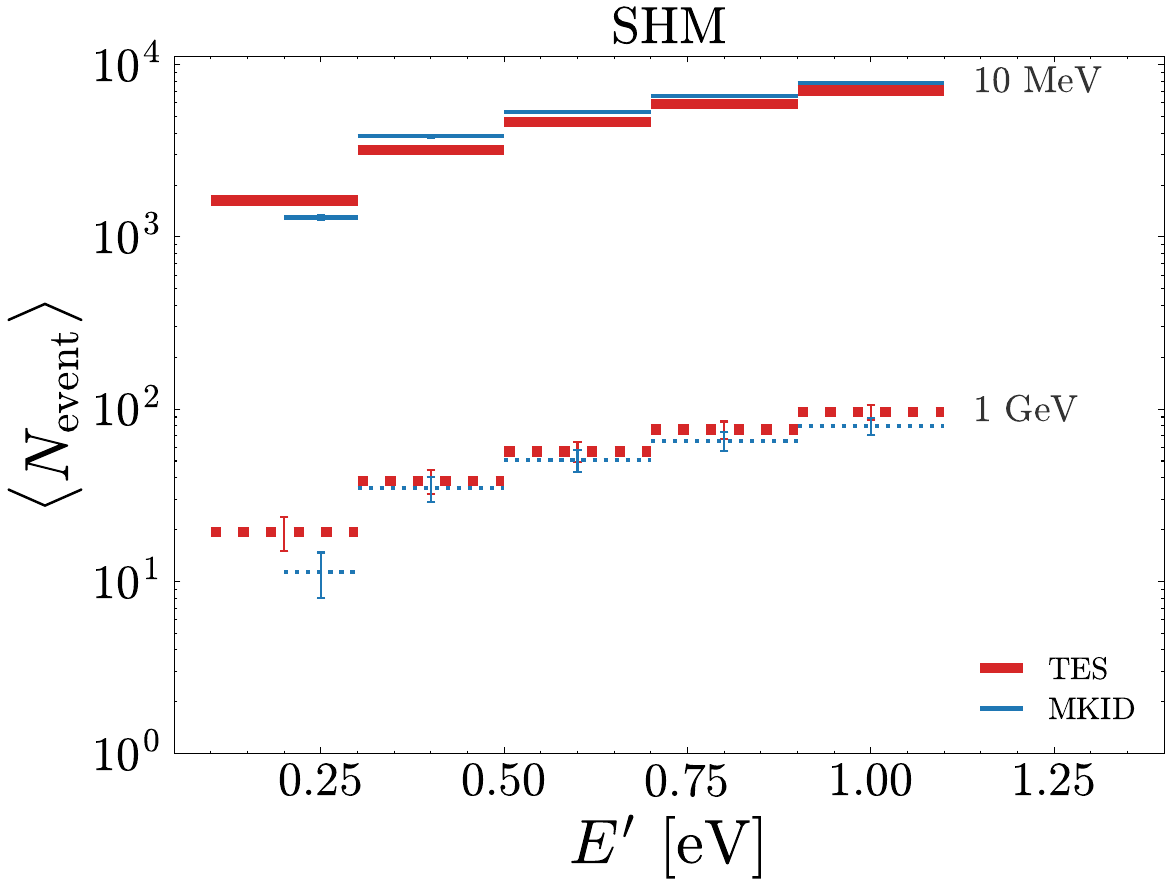}
    \includegraphics[width=0.32\linewidth]{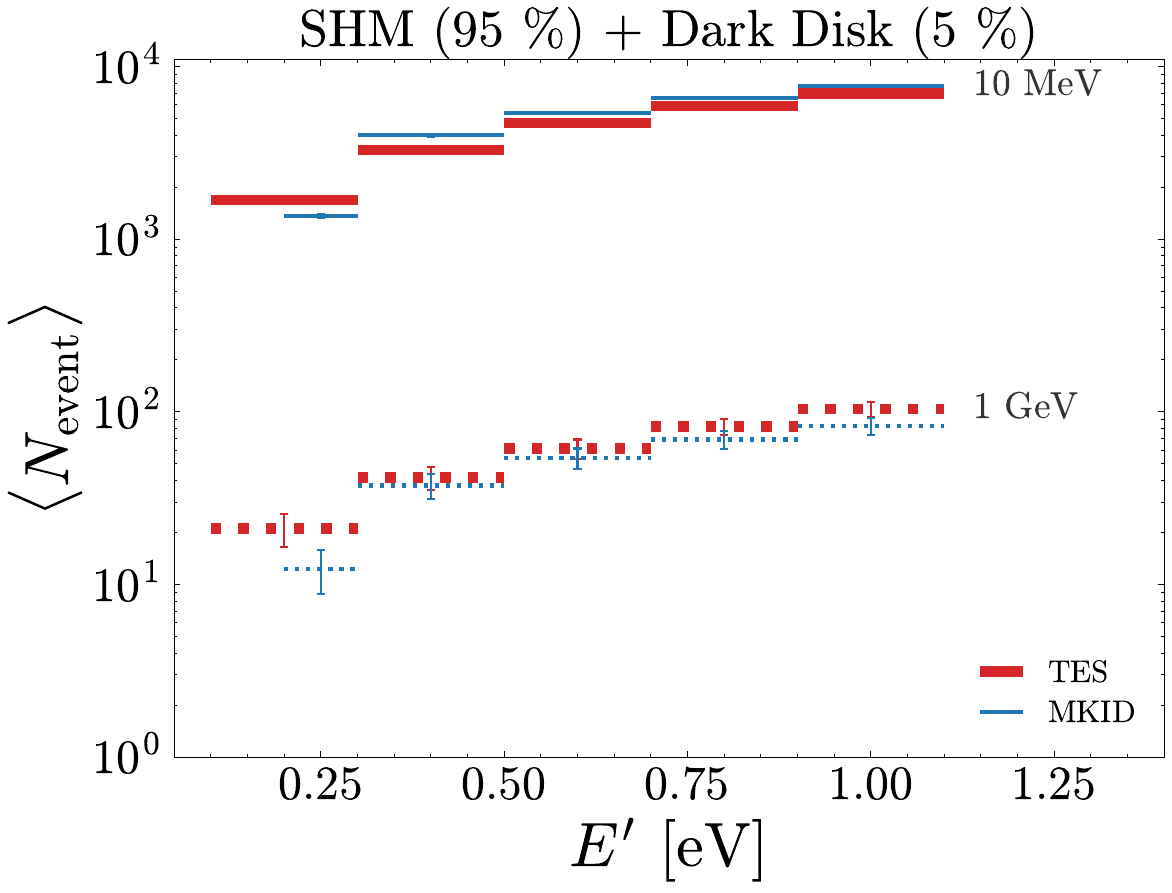}
    \includegraphics[width=0.32\linewidth]{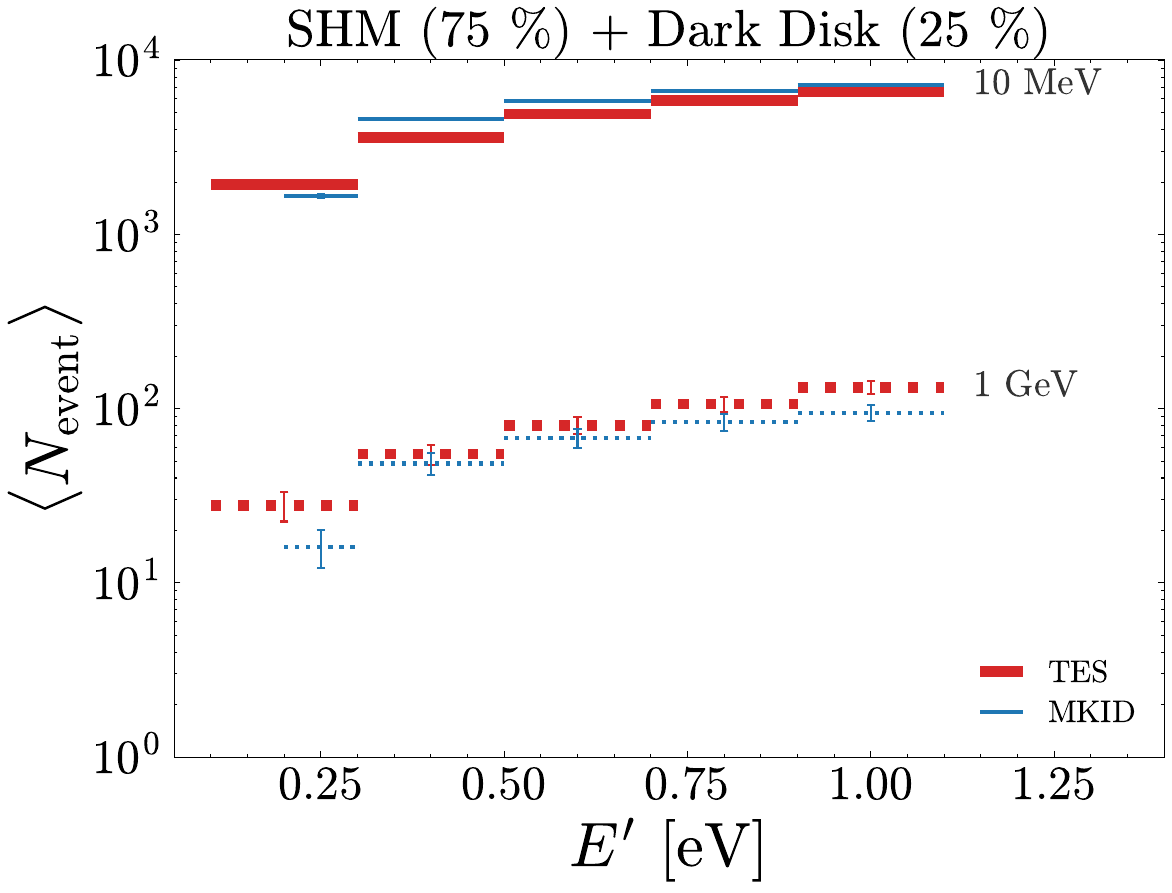}
    \caption{Predicted event counts per observed energy bin  $\langle N_{\rm event}\rangle_i$  (that we take as mock data) for the TES (orange) and MKID (blue) configurations, for the  SHM (Left),  SHM$(95\%)+{\rm DD}(5\%)$ (Middle) and SHM$(75\%)+{\rm DD}(25\%)$ (Right) benchmark halo models. Solid and dotted horizontal segments denote $m_\chi=10~\MeV$ and $m_\chi=1~\GeV$ DM masses, respectively. The error bars indicate statistical (Poisson) uncertainties.}
    \label{eventNumbers}
\end{figure*}

Following Ref.~\cite{Peter:2011eu} we model this scenario as a truncated Maxwellian of the
form of Eq.~\eqref{SHM_dif} with values
$\rho_\tx{EB} = 10^{14}~\GeV$~\cm$^{-3}$, $v_0^\tx{EB} = \sqrt{2k_BT/m_\chi}$, $v_e^\tx{EB} = 0$,
$v_\tx{esc}^\tx{EB} = 11.2~$km~s$^{-1}$~\cite{Gould:1987ir, Neufeld:2018slx,
McKeen:2022poo}. Here, $T \simeq 300~\tx{K}$ is the ambient terrestrial temperature, so that
$v_0^\tx{EB}$ is the thermal speed in equipartition with terrestrial
matter, so that $v_0^\tx{EB} = 2.16~$km~s$^{-1}$ for $m_\chi = 1~\GeV$, and $v_e^\tx{EB} = 0$
encodes the co-moving character of a bound component, and $v_\tx{esc}^\tx{EB}$ is
Earth's surface escape speed. The total halo function is additive,
\al{
\wt\eta_\tx{SHM+EB}(v_\min) =
\wt\eta_\tx{SHM}(v_\min) + \wt\eta_\tx{EB}(v_\min),
\label{etaSHMpEB}
}
each term carrying its own density prefactor in Eq.~\eqref{halofuncDef}
with $\rho_\tx{DM}$ for the SHM and $\rho_\tx{EB}$ for the EB. Since $\rho_\tx{EB}/\rho_\tx{DM} \simeq 10^{14}$, the EB term dominates
$\wt\eta_\tx{SHM+EB}$ throughout the accessible region
$v_\min \lesssim v_\tx{esc}^\tx{EB}$.

\subsection{Mock Data Event Rates}\label{data}

Although the detector configurations considered here are sensitive to individual events, reconstructing the halo function benefits from spectral information. We therefore analyze the distribution of assumed signal events in bins of observed energy. For concreteness, for both detector configurations we analyze we bin the observed energy $E'$ into five bins spanning from
each detector's threshold up to $1.1~\mathrm{eV}$. The upper value is chosen to
concentrate on the sub-eV to eV window enabled by low threshold quantum sensors,
which is where these detectors provide reach beyond conventional direct DM detection
experiments and which maps onto the low $v_\min$ region we aim to reconstruct. 
For the TES configuration, with a threshold of
$0.1~\mathrm{eV}$, these are five uniform bins of width $0.2~\mathrm{eV}$ giving
$[0.1,0.3]$, $[0.3,0.5]$, $[0.5,0.7]$, $[0.7,0.9]$ and $[0.9,1.1]~\mathrm{eV}$.
For the MKID configuration, whose threshold is $0.2~\mathrm{eV}$, we keep the same upper four
bins and use a narrower lowest bin $[0.2,0.3]~\mathrm{eV}$, giving
$[0.2,0.3]$, $[0.3,0.5]$, $[0.5,0.7]$, $[0.7,0.9]$ and $[0.9,1.1]~\mathrm{eV}$.

In the $i$-th observed energy bin $[E_i^\pr, E_{i+1}^\pr]$, the  event rate is
\al{
    R_i = R_{[E_i^\pr, E_{i+1}^\pr]}
    = \int_{E_i^\pr}^{E_{i+1}^\pr}\df E^\pr 
      {\df R\ov \df E^\pr}\fn{E^\pr},
}
and the expected predicted number of events in that bin is
\al{
    \braket{N_\tx{event}}_i = M_T T R_i,
    \label{eventNumber}
}
where $M_T$ is the detector mass and $T$ the total running time. 
For the projected exposures, we adopt
\al{
M_T T = \begin{cases}
8.2~\mathrm{\mu g\cdot month} & (\tx{TES}~\cite{Chen:2025cvl}),\\[2pt]
10^7~\mathrm{pixel\cdot year}  = 4.2~\mathrm{m g\cdot year} & (\tx{MKID}~\cite{Gao:2024irf}),
\end{cases}
\label{exposures}
}
where a target mass of $0.42~\tx{ng}$ per pixel is considered for the
MKID configuration~\cite{2024SuScT..37e5014H}. The computed expected event numbers for all three
benchmark models at these exposures are shown in Figs.~\ref{eventNumbers}
and~\ref{Bound}.

\subsection{Reconstruction of Speed Distributions}\label{results}

\begin{figure*}[t]
    \centering
  \includegraphics[width=0.45\linewidth]{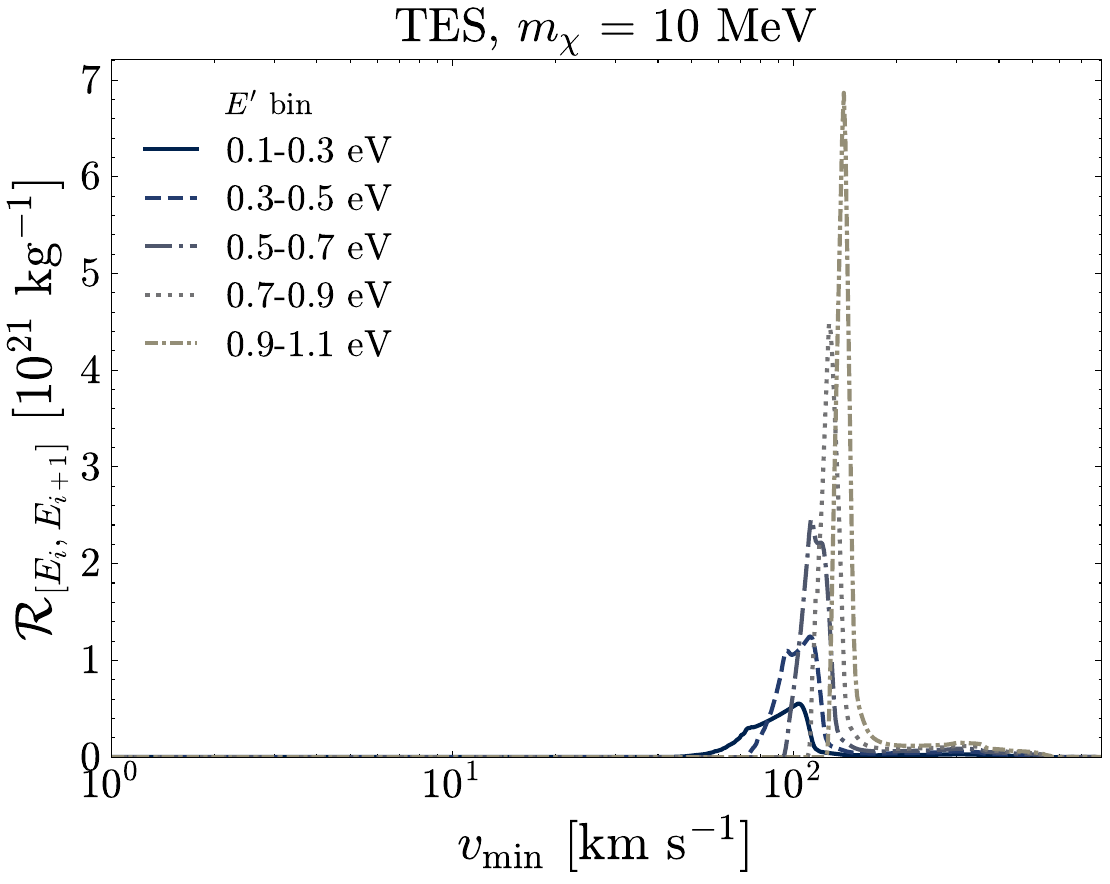}
\includegraphics[width=0.45\linewidth]{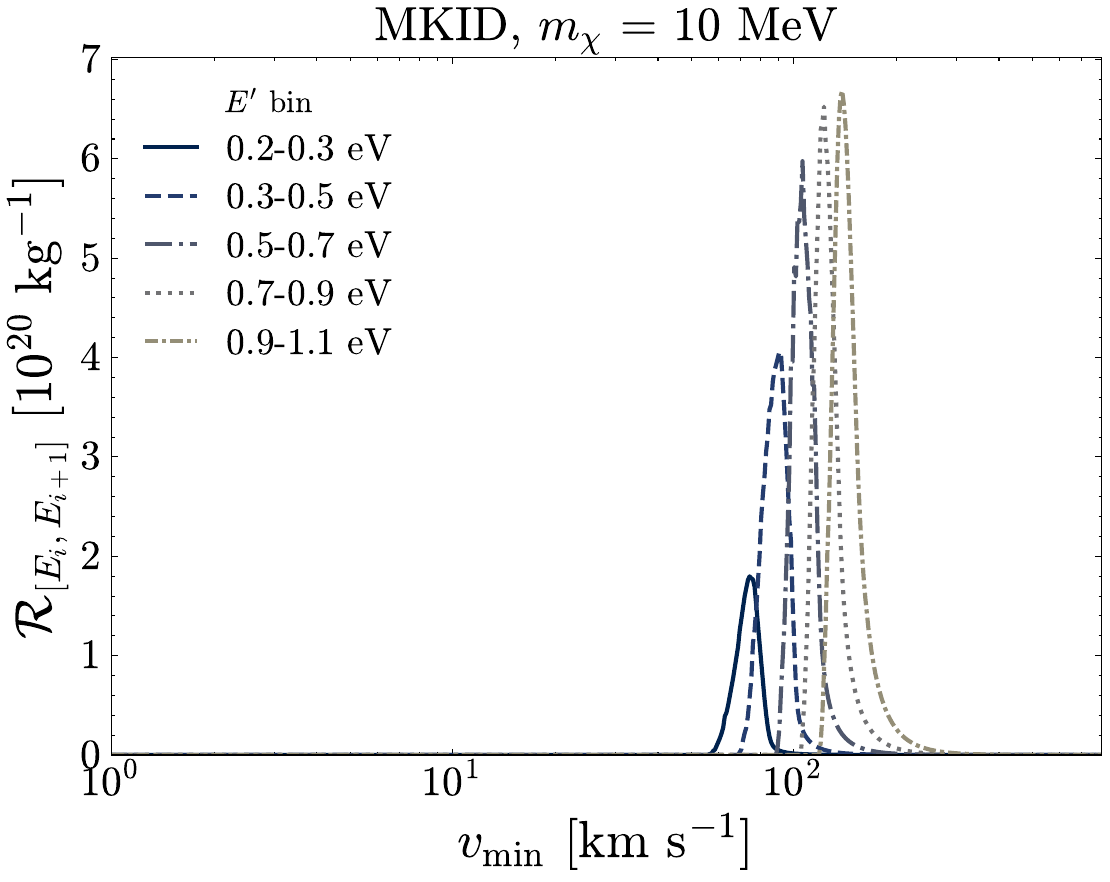}
   \includegraphics[width=0.45\linewidth]{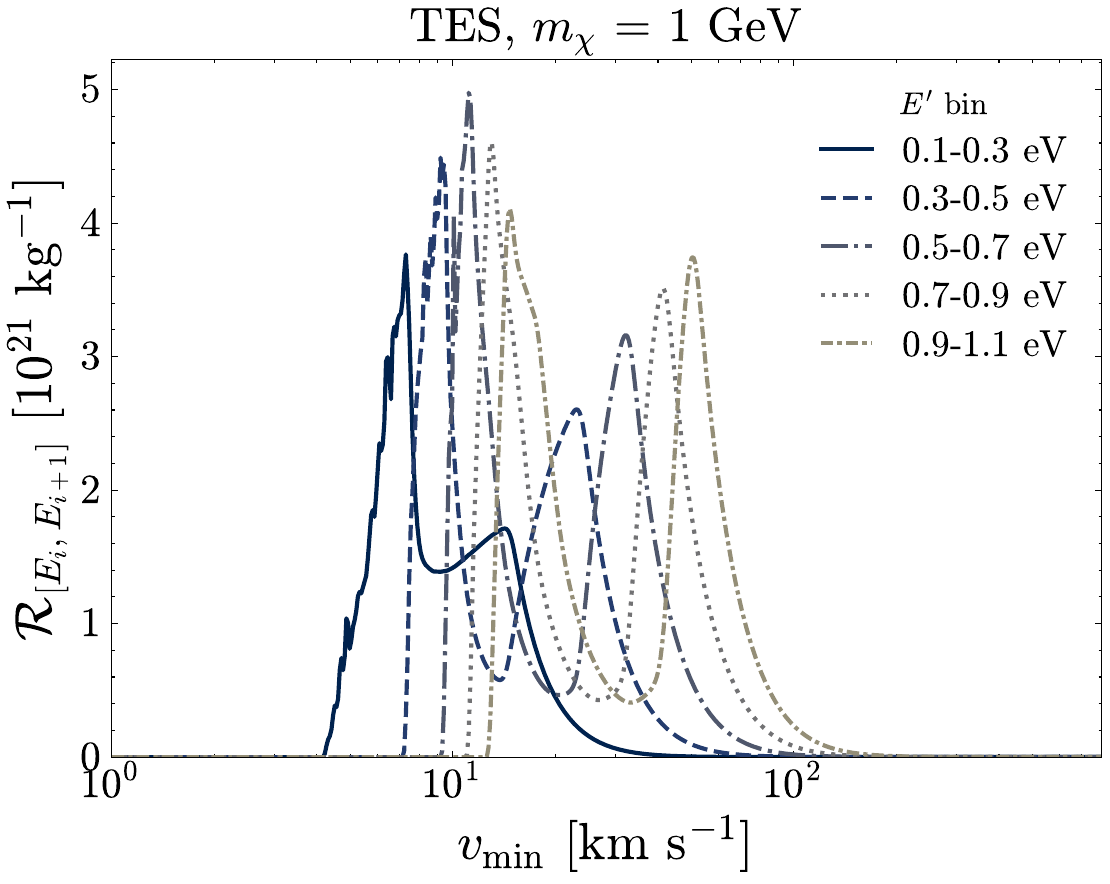}
   \includegraphics[width=0.45\linewidth]{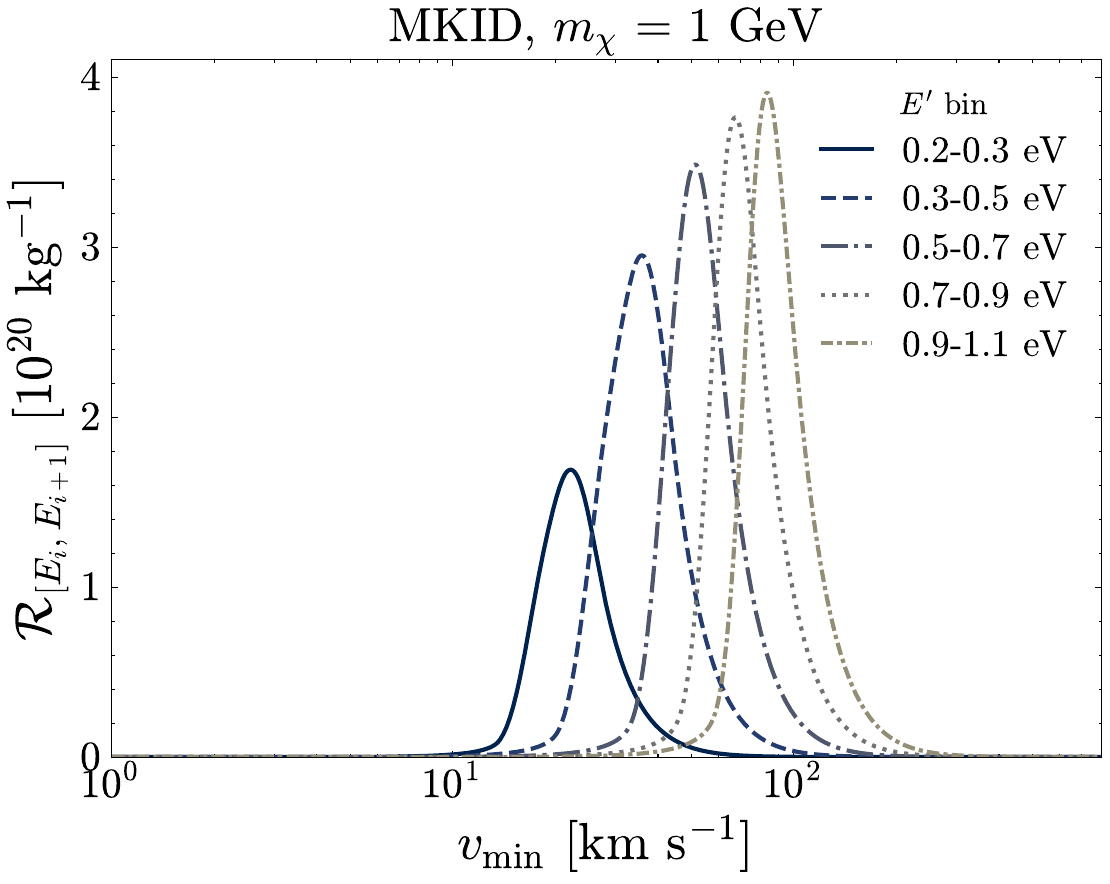}
\caption{Integrated response functions $\mathcal{R}_{[E'_i,E'_{i+1}]}(v_{\min})$ for DM-electron scattering with a heavy mediator, shown as functions of $v_{\min}$. The left and right panels correspond to the TES and MKID configurations, respectively, while the upper and lower rows correspond to $m_\chi=10~\MeV$ and $m_\chi=1~\GeV$. Each curve corresponds to one observed energy interval $[E'_i,E'_{i+1}]$ as indicated. }
    \label{curlyR}
\end{figure*}

The event rate in the $i$-th observed energy bin can be stated as a convolution
of the halo function with the response integrated over that bin,
\al{
    R_i = \int \df v_\min
    \mc R_{\sqbr{E_i^\pr, E_{i+1}^\pr}}\!\fn{v_\min}
    \wt\eta\fn{v_\min},
}
where the integrated response function is
\al{
    \mc R_{\sqbr{E_i^\pr, E_{i+1}^\pr}}\!\fn{v_\min}
    = \int_{E_i^\pr}^{E_{i+1}^\pr}\df E^\pr
    {\df \mc R\ov \df E^\pr}\fn{v_\min, E^\pr}.
    \label{binCurlyR}
}
This function is shown in Fig.~\ref{curlyR} for the TES and MKID benchmarks over
several $E^\pr$ bins. A given bin constrains $\wt\eta(v_\min)$ only over the range
of $v_\min$ where its $\mc R_{\sqbr{E_i^\pr, E_{i+1}^\pr}}$ is appreciably nonzero.
Thus, the integrated responses effectively act as window functions in $v_\min$.

\begin{figure}[t]
    \centering
\begin{tikzpicture}[scale = 0.8]
  \draw[-stealth] (-0.5,0) -- (7.5,0) node[right] {$v_\min$};
  \draw[-stealth] (-0.5,0) -- (-0.5,4.5) node[above] {$\wt \eta_{\rm ansatz}(v_\min)$};

  \def\width{0.7}

  \def\hA{3.5}
  \def\hB{3.4}
  \def\hC{2.8}
  \def\hD{2.0}
  \def\hJ{1.5}
  \def\hN{1.0}

  \draw[thick,black] (0,0) -- (0,\hA);
  \draw[thick,black] (0,\hA) -- (\width,\hA) node[midway, above]{$\wt\eta_1$};
  \draw[thick,black] (\width,\hA) -- (\width,0) node[below] {$v_\min^2$};
  \draw[black] (0,0) -- (\width,0) node[midway, below left, black] 
  {$v_\min^1$};

  \draw[thick,black] (\width,\hB) -- (2*\width,\hB) node[midway, above] {$\wt\eta_2$};
  \draw[thick,black] (2*\width,\hB) -- (2*\width,0) node[below] {$v_\min^3$};

  \draw[thick,black] (2*\width,\hC) -- (3*\width,\hC) node[midway, above] {$\wt\eta_3$};
  \draw[thick,black] (3*\width,\hC) -- (3*\width,0) node[below] {$v_\min^4$};

  \draw[thick,black] (3*\width,\hD) -- (4*\width,\hD) node[midway, above] {$\wt\eta_4$};
  \draw[thick,black] (4*\width,\hD) -- (4*\width,0) node[below] {$v_\min^5$};

  \fill (4.5*\width, 0.8) circle (1pt);
  \fill (4.7*\width, 0.8) circle (1pt);
  \fill (4.9*\width, 0.8) circle (1pt);

  \fill (6.8*\width, 0.8) circle (1pt);
  \fill (7.0*\width, 0.8) circle (1pt);
  \fill (7.2*\width, 0.8) circle (1pt);

  \draw[thick,black] (7.6*\width,0) node[below] {$v_\min^{N_\tx{int}}$} -- (7.6*\width,\hN);
  \draw[thick,black] (7.6*\width,\hN) -- (8.6*\width,\hN) node[midway, above] {$\wt\eta_{N_\tx{int}}$};
  \draw[thick,black] (8.6*\width,\hN) -- (8.6*\width,0) node[below right] {$v_\min^{N_\tx{int} + 1}$};

  \draw[thick,black] (5.3*\width,0) node[below] {$v_\min^j$} -- (5.3*\width,\hJ);
  \draw[thick,black] (5.3*\width,\hJ) -- (6.3*\width,\hJ) node[midway, above] {$\wt\eta_j$};
  \draw[thick,black] (6.3*\width,\hJ) -- (6.3*\width,0);

\end{tikzpicture}
    \caption{ Schematic discretization of the halo function $\widetilde{\eta}(v_{\min})$ into $N_{\rm int}$ intervals, with one value $\widetilde{\eta}_j$ assigned to each interval. The $\widetilde{\eta}_j$ are the parameters that we fit. Under the positivity and monotonicity restrictions many adjacent parameters  take the same value leading to a solution maximizing the likelihood that is
    a piecewise-constant function with at most $(d-1)$ downward steps, where $d$ is the number of data entries (here, the number of energy bins).
    }
    \label{haloFunctions-hist} 
\end{figure}
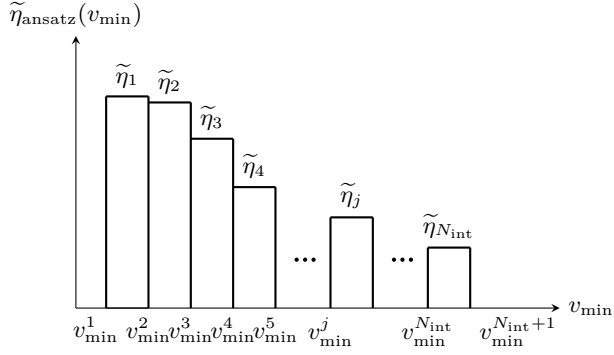

Following the procedure of Ref.~\cite{Feldstein:2014gza}, we discretize the halo
function on the $v_\min$ interval of interest, namely in the range over which the response
functions are appreciably nonzero as shown in Fig.~\ref{curlyR}. We divide it into
$N_\tx{int}$ equal adjacent intervals $j = 1,\dots,N_\tx{int}$ on each of which
$\wt\eta$ takes a unique value $\wt\eta_j$,
\al{
    \wt\eta_{\rm ansatz}(v_\min)
    = \sum_{j=1}^{N_\tx{int}} \wt\eta_j
    \sqbr{\Theta\fn{v_\min - v_\min^j}
    - \Theta\fn{v_\min - v_\min^{j+1}}},
    \label{etaStepForm}
}
as illustrated in Fig.~\ref{haloFunctions-hist}. The $\wt\eta_j$ are the parameters
fitted by the constrained minimization described below. In the best-fit, multiple
adjacent intervals take the same value. Hence, the reconstructed $\wt\eta(v_\min)$ is
piecewise constant with at most $(d-1)$ downward steps, where $d$ is the number of data bins. The interval boundaries serve to locate these steps, since each step must fall on one of them. We therefore use a grid much finer than the expected number of steps
and verify convergence to a particular best-fit function. Namely starting from a number of intervals larger than the number
of data bins and doubling it, the best fit shape ceases to change once the grid is
sufficiently fine to resolve the step locations, after which further refinement leaves
them fixed.

Inserting the ansatz Eq.~\eqref{etaStepForm} into the rate, the predicted
differential rate is
\al{
    {\df R\ov \df E^\pr}\fn{E^\pr}
    = \sum_{j=1}^{N_\tx{int}} \wt\eta_j
    \int_{v_\min^j}^{v_\min^{j+1}}\df v_\min
    {\df \mc R\ov \df E^\pr}\fn{v_\min, E^\pr}.
}
Its integral over bin $i$ is
\al{
    R_i =&~ \sum_{j=1}^{N_\tx{int}} \wt\eta_j
    \int_{E_i^\pr}^{E_{i+1}^\pr}\df E^\pr
    \int_{v_\min^j}^{v_\min^{j+1}}\df v_\min
    {\df \mc R\ov \df E^\pr}\fn{v_\min, E^\pr} \nn
    =&~ \sum_{j=1}^{N_\tx{int}}   \mc R_{ij}\wt\eta_j,
}
which defines the response matrix elements
\al{
    \mc R_{ij}
    =&~ \int_{E_i^\pr}^{E_{i+1}^\pr}\df E^\pr
    \int_{v_\min^j}^{v_\min^{j+1}}\df v_\min
    {\df \mc R\ov \df E^\pr}\fn{v_\min, E^\pr} \nn
    =&~ \int_{v_\min^j}^{v_\min^{j+1}}\df v_\min
    \mc R_{\sqbr{E_i^\pr, E_{i+1}^\pr}}\!\fn{v_\min},
    \label{responseMatrix}
}
with the second form following from Eq.~\eqref{binCurlyR}.

As mock observed data $\mc O_i$ we take the per bin event numbers
$\mc O_i = \braket{N_\tx{event}}_i$ of Eq.~\eqref{eventNumber}, computed for each of
the three benchmark DM velocity distributions. We obtain the best fit
$\wt\eta(v_\min)$ by minimizing the Neyman $\chi^2$ statistic,
\al{
    \chi^2_\tx{N} = \sum_i \frac{(\mu_i - \mc O_i)^2}{\mc O_i},
    \label{chiSq}
}
in which the variance of each bin is approximated by its observed Poisson count
$\mc O_i$, so that the statistical uncertainty is
$\sigma_{\tx{stat},i} \simeq \sqrt{\mc O_i}$. This Gaussian approximation to the
Poisson likelihood is accurate for bins with about ten events or more. It is one
convenient choice and not a strict requirement of the HI method, and other
statistics such as the Poisson likelihood $\chi^2$ of Ref.~\cite{Baker:1983tu}
could be used instead. In this simplified analysis we neglect backgrounds, such as
a possible low energy excess~\cite{Anthony-Petersen:2024vdh}, as well as systematic
uncertainties.  

The
predicted event number per bin is
\al{
    \mu_i = M_T T \sum_{j=1}^{N_\tx{int}} \mc R_{ij} \wt\eta_j,
    \label{mudef}
}
and the minimization procedure determines $\wt\eta_j$, yielding  an approximation to the input $\wt{\eta}$ function. 
During the minimization we impose two physical constraints,
\al{
    \wt\eta_j \geq 0,
    \qquad
    \wt\eta_j \geq \wt\eta_{j+1}
    \quad (j = 1,\dots,N_\tx{int}-1),
}
enforcing the positivity of the DM speed distribution and the non-increasing
character of any physical halo integral.

The statistical Poisson uncertainties on the input counts $\mc O_i$ are shown in
Fig.~\ref{eventNumbers}. Propagating them through the constrained minimization
requires Monte Carlo methods and is deferred to future work.

We note that since our likelihood is based on binned event counts, as explained in Sec.~\ref{sec:HIframework}, the best fit halo function need not be unique~\cite{Gelmini:2017aqe}. The minimization therefore returns a piecewise constant best-fit solution. Other halo functions may yield the same minimum value of the test statistic and identical predicted bin counts. We do not determine the corresponding degeneracy or confidence bands here.

\begin{figure*}[t]
    \centering
    \includegraphics[width=0.45\linewidth]{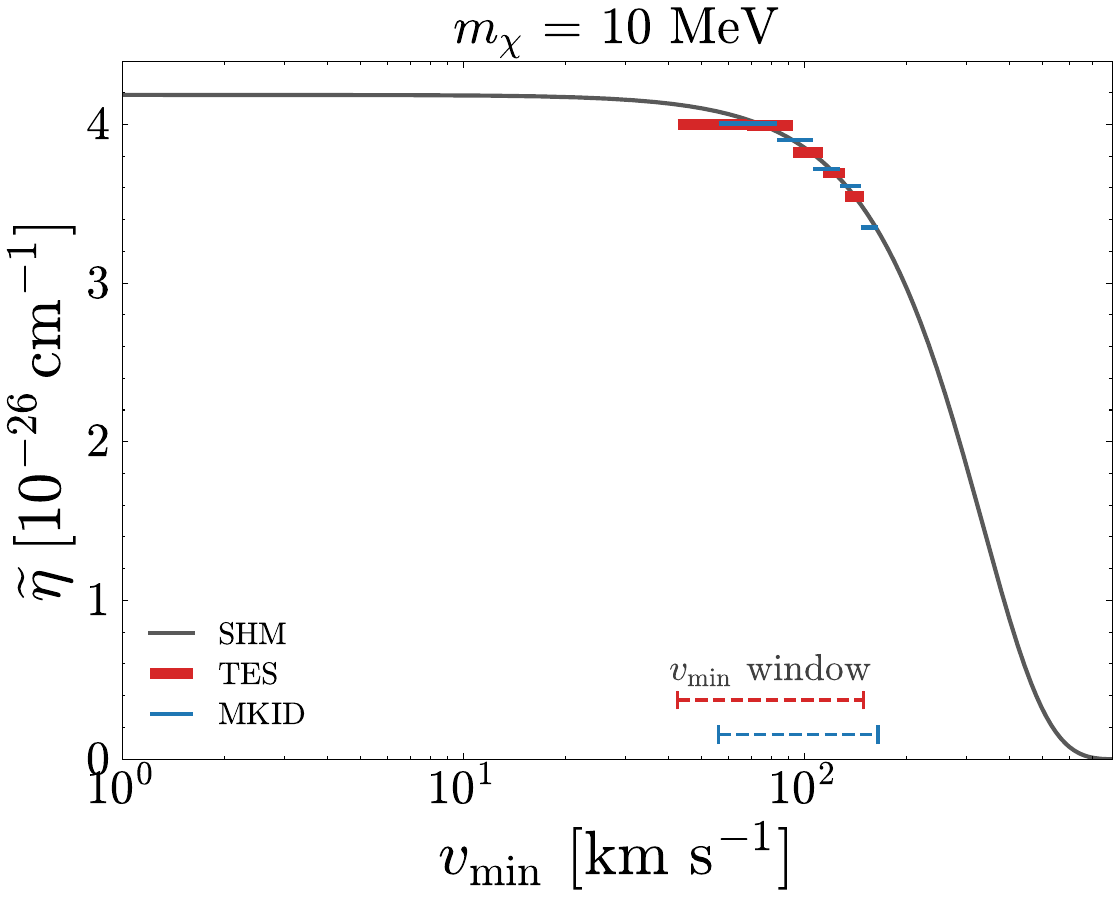}
    \includegraphics[width=0.45\linewidth]{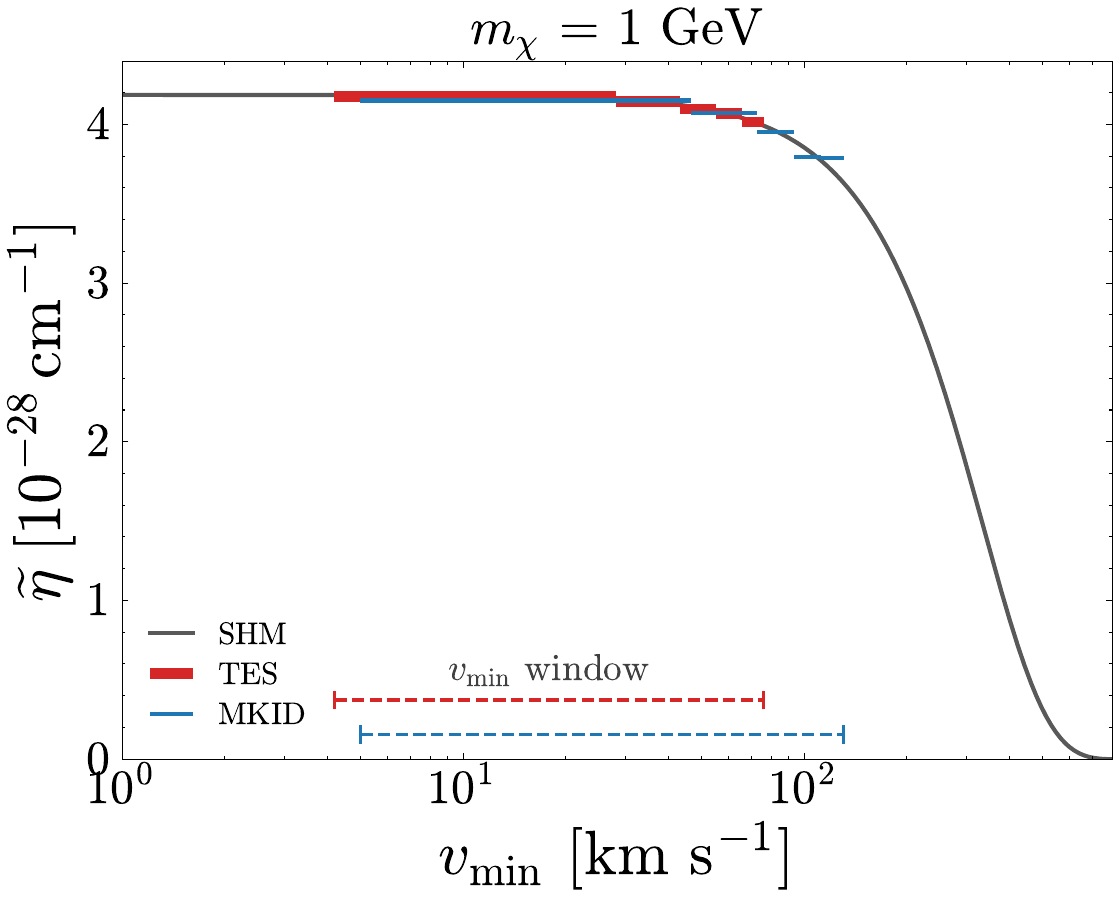}
    \caption{Reconstruction of the halo function $\wt{\eta}(v_\min)$ for the SHM benchmark. The black curve shows the input SHM halo function, while the horizontal segments show the best fit piecewise-constant reconstructions obtained separately from the TES (orange) and MKID (blue) mock event counts. The left and right panels correspond to $m_\chi=10~\MeV$ and $m_\chi=1~\GeV$, respectively. The dashed horizontal lines specify the $v_\min$ range for which each experiment has sensitivity, namely where their respective response function depart significantly from zero.
    }
    \label{SHM}  
\end{figure*}

\begin{figure*}[t]
    \centering
    \includegraphics[width=0.43\linewidth]{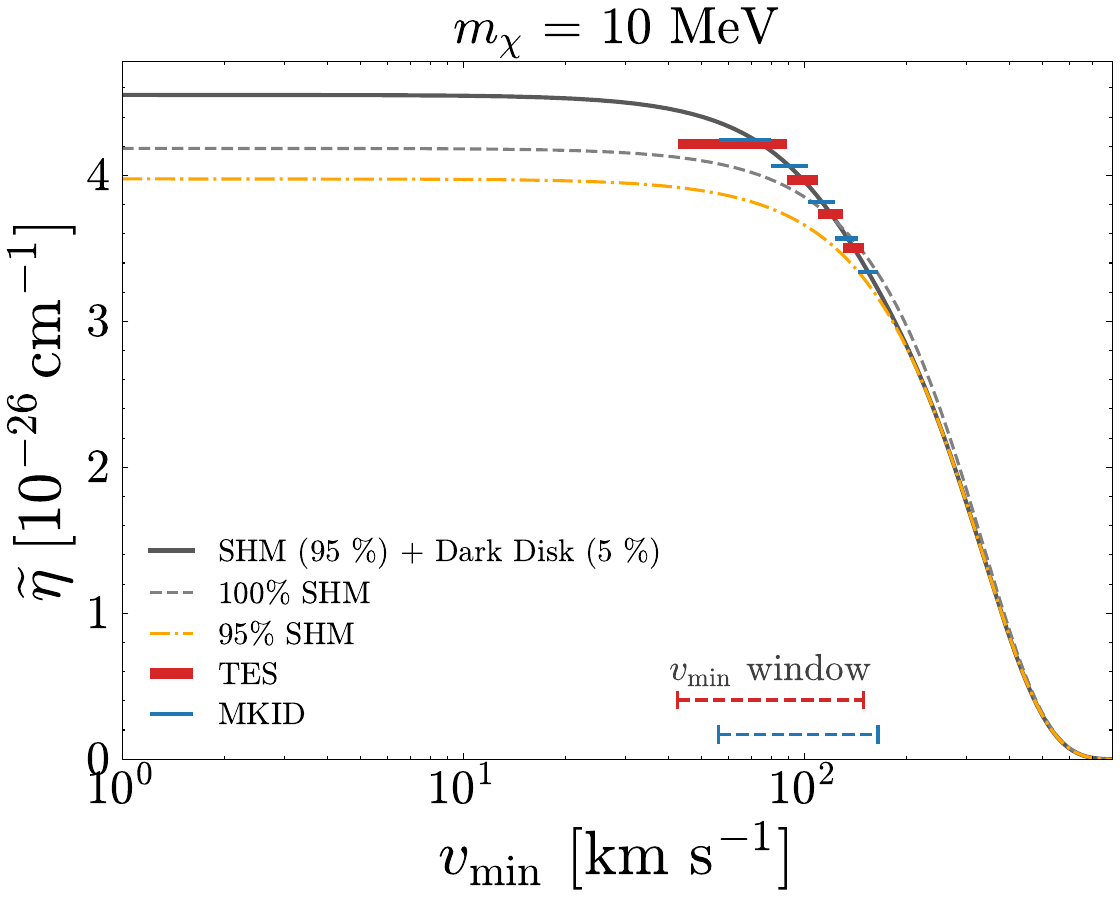}
    \includegraphics[width=0.43\linewidth]{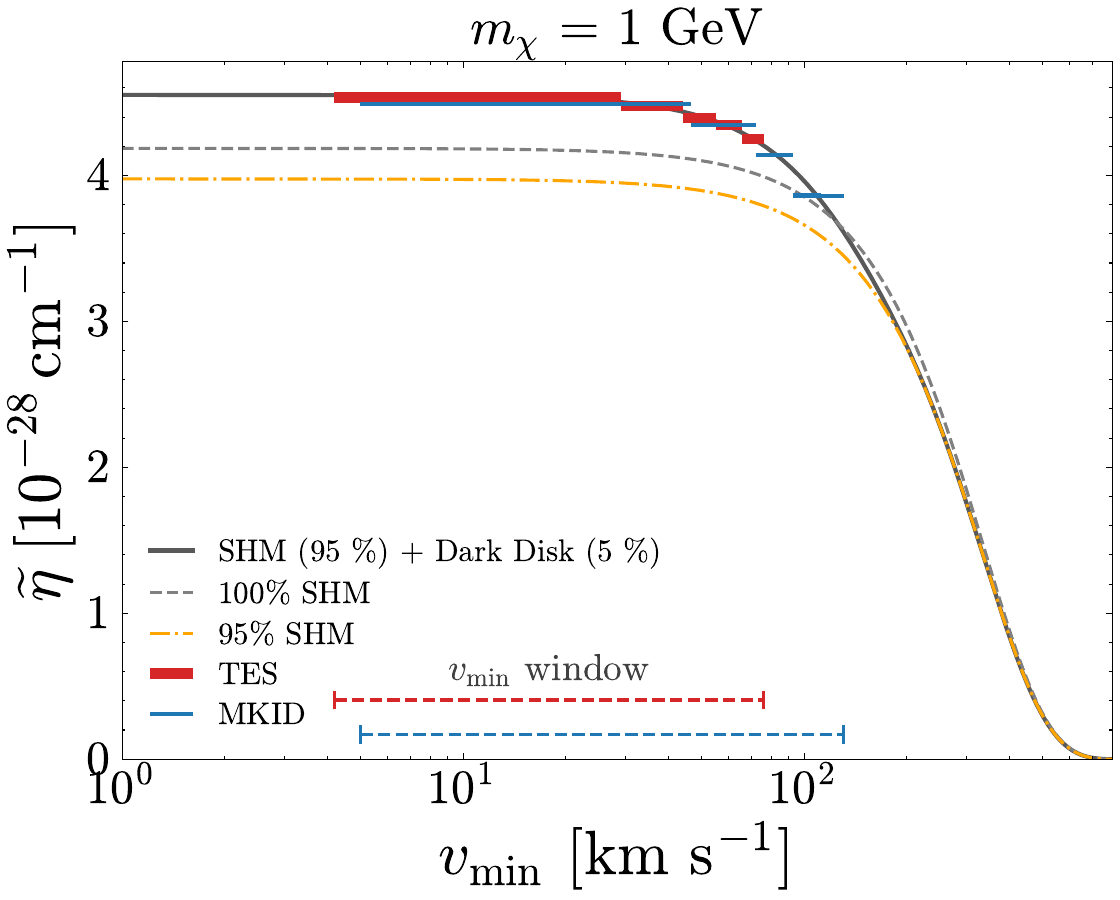}

    \includegraphics[width=0.43\linewidth]{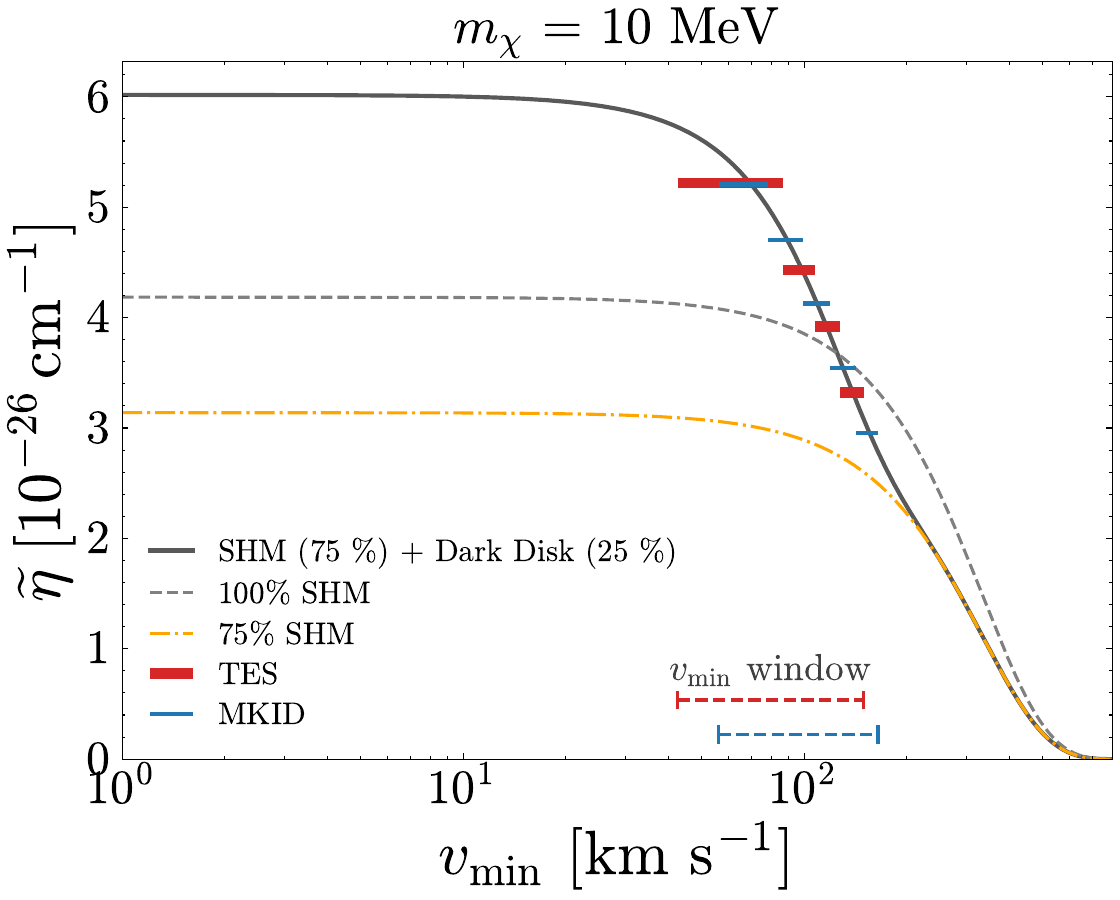}
    \includegraphics[width=0.43\linewidth]{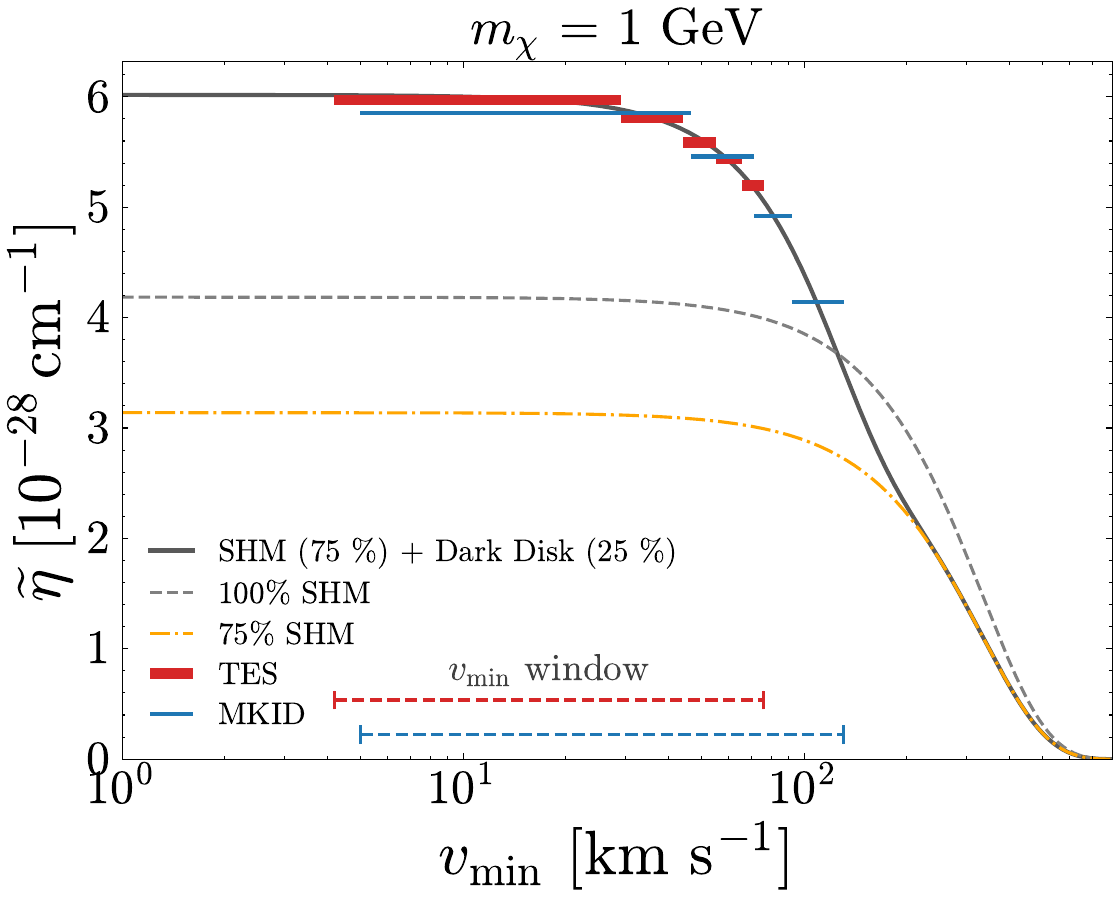}
    \caption{Reconstruction of the halo function $\wt{\eta}(v_\min)$ for the SHM+DD benchmarks. The upper and lower rows correspond to dark-disk fractions $f_\tx{DD}=5\%$ and $25\%$, respectively, while the left and right columns correspond to $m_\chi=10~\MeV$ and $m_\chi=1~\GeV$. In each panel the solid black curve shows the full input SHM+DD halo function, the dashed black curve shows the pure SHM result at the same total local density  and the dash-dotted black curve shows the SHM contribution weighted by $(1-f_\tx{DD})$. The horizontal segments show the best-fit piecewise-constant reconstructions obtained separately from the TES (orange) and MKID (blue) mock event counts. The dashed horizontal lines specify the $v_\min$ range for which the experiments have sensitivity, namely where their response function depart significantly from zero.
    }
    \label{DarkDisk} 
\end{figure*}

\subsection{Reconstructed Halo Functions}

\begin{figure*}[t]
    \centering
    \includegraphics[width=0.45\linewidth]{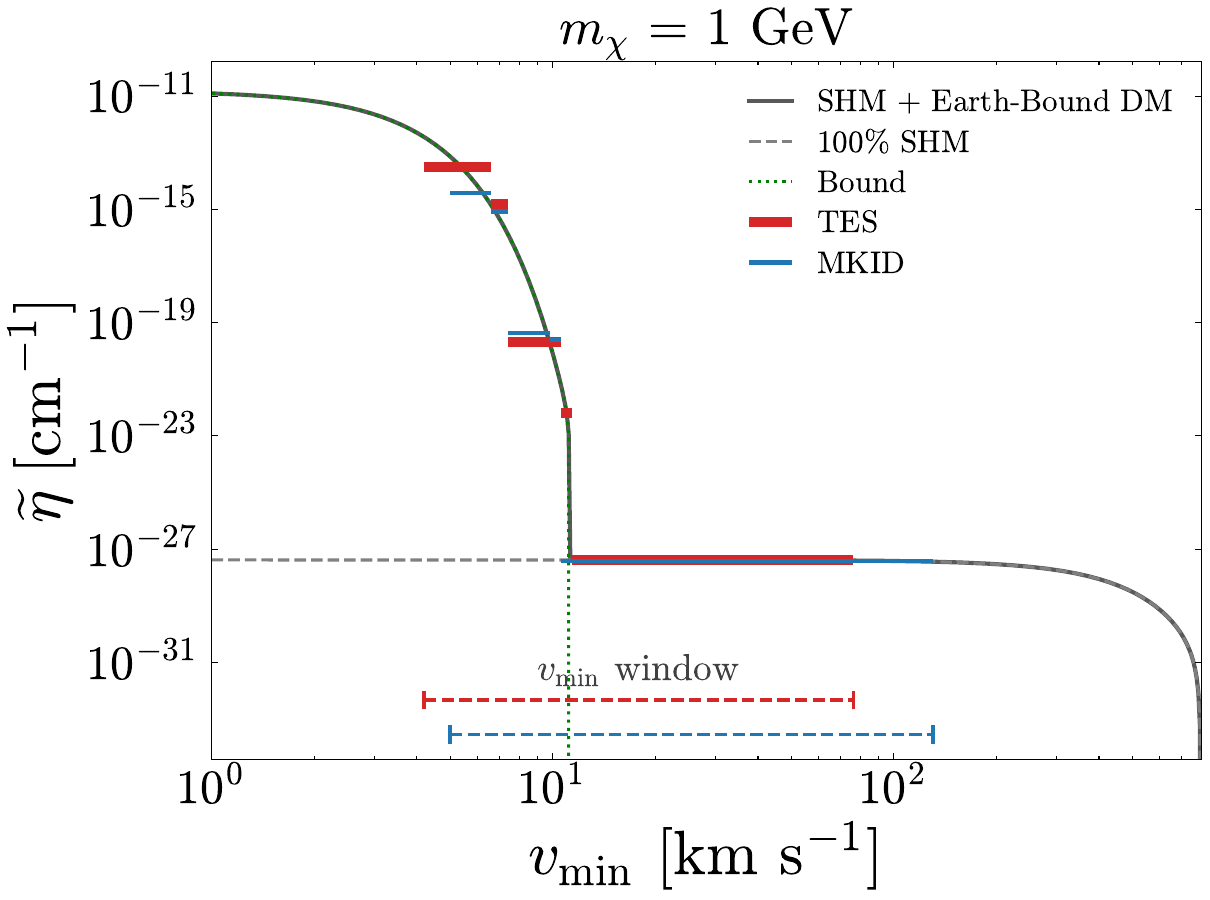}
    \includegraphics[width=0.433\linewidth]{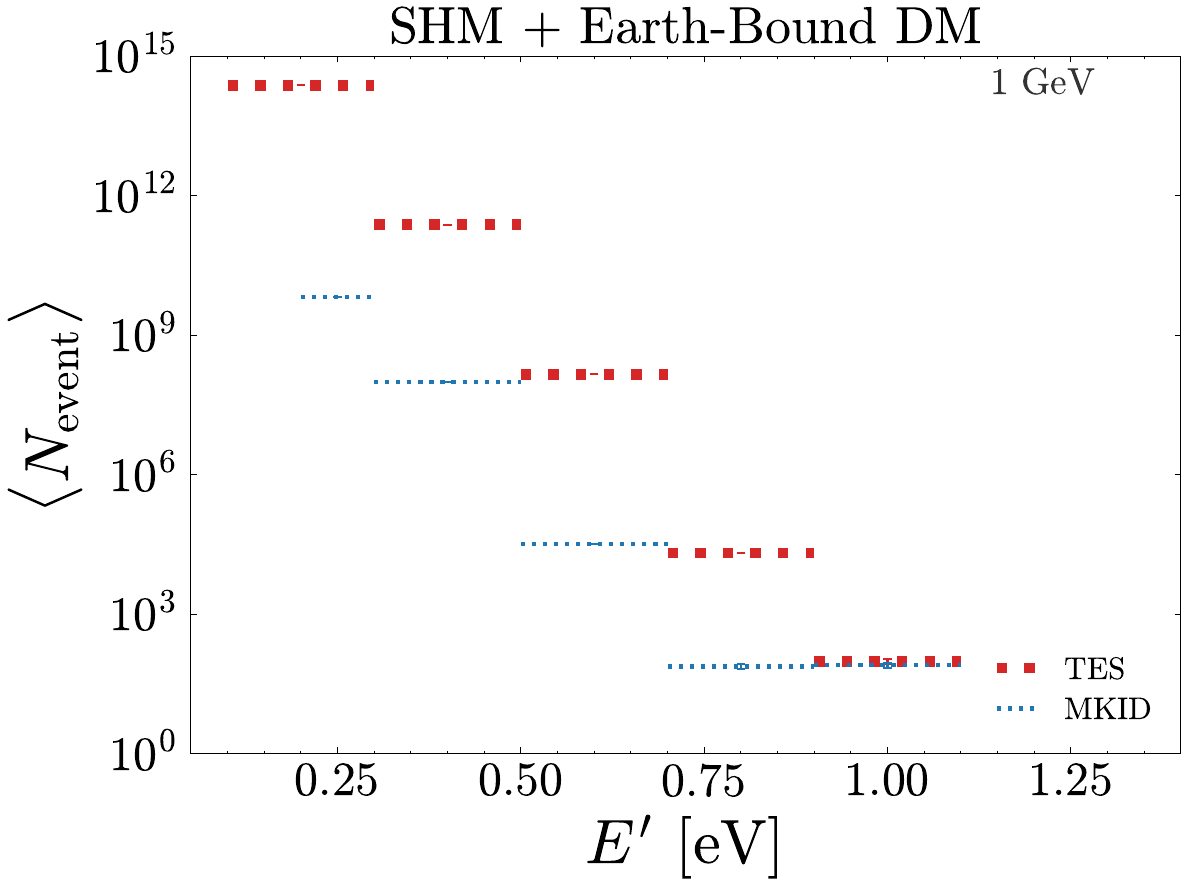}
    \caption{(Left) Reconstruction of the halo function $\wt{\eta}(v_\min)$ for the SHM+EB benchmark with $m_\chi=1~\GeV$. The solid black curve shows the full input SHM+EB halo function  while the dashed and dotted black curves show the SHM and Earth-bound contributions, respectively. The horizontal segments show the best fit piecewise-constant reconstructions obtained separately from the TES (orange) and MKID (blue) mock event counts. The dashed horizontal lines specify the $v_\min$ range for which the experiments have sensitivity, namely where their response function depart significantly from zero. 
    (Right) Expected event counts per observed energy bin  $\langle N_\tx{event}\rangle_i$ for the same benchmark and detector configurations, which we take as our observed mock data.  The error bars indicate statistical (Poisson) uncertainties.}
    \label{Bound} 
\end{figure*}

In Figs.~\ref{SHM} \ref{DarkDisk} and \ref{Bound} we show  the original and best-fit $\wt\eta(v_\min)$ functions, for  TES
and MKID configurations, for $m_\chi = 10$~MeV  and 1 GeV.

In the analysis, for each mass and
detector configuration the $v_\min$ axis is divided into $N_{\rm int}$  intervals covering the kinematically accessible window. Concretely, for the TES (Al) and $m_\chi= 10$ MeV, we divide the window $\sqbr{42.5~\km~\s^{-1}, 150~\km~\s^{-1}}$ into $135$ intervals, and for $m_\chi= 1$ GeV, we divide  $\sqbr{4.2~\km~\s^{-1}, 78~\km~\s^{-1}}$ into $92$ intervals (see in Fig.~\ref{curlyR} where the response functions are non-zero). For the MKID (TiN), we divide $\sqbr{56.1~\km~\s^{-1}, 168~\km~\s^{-1}}$ into $140$ intervals when $10\ \mathrm{MeV}$ and $\sqbr{5.0~\km~\s^{-1}, 134~\km~\s^{-1}}$ into $162$ intervals when $1\ \mathrm{GeV}$. In every case the intervals are equally spaced, with width $0.8~\km~\s^{-1}$.
Since the spacing is linear while the features that
distinguish the benchmark models lie at low $v_\min$, a small fraction of the
linear range but a broad region on the logarithmic $v_\min$ axis of
Figs.~\ref{SHM}, \ref{DarkDisk} and~\ref{Bound}, a large $N_\tx{int}$ is required to
position sufficient  interval boundaries at low velocities.
The height of the horizontal bars indicate constant pieces of the best-fit, which are at most five, and their projection into the horizontal axis the $v_\min$ interval they correspond to. 

In Fig.~\ref{SHM}, for $m_\chi = 10~\MeV$, the $v_\min$ ranges to which both  detector configurations are sensitive correspond to the $\wt\eta_{\rm SHM}$ slowly varying region, below where
$\wt\eta_\tx{SHM}$ falls near $v_\min \simeq 294~$km~s$^{-1}$. Their
reconstructions closely trace the input. TES and MKID cover offset although
partially overlapping $v_\min$ windows, determined by their different material responses
(i.e. Al versus TiN) and resolutions. In the overlapping range the two best fits agree with each
other and with the SHM, which serves as an internal consistency check of the method  since a
disagreement there could point to   detector systematic effects or an incorrect response
model rather than to a real feature of $\wt\eta$. For $m_\chi = 1~\GeV$ the
kinematic minimum $v_{\ast}$ is smaller by a factor
$\mathcal{O}(10)$, and both detectors reach $\mc O(10)~$km~s$^{-1}$. There $\wt\eta_\tx{SHM}$ is
essentially constant, thus the reconstruction primarily fixes the normalization factor
$\rho_\tx{DM} \sigma_e/m_\chi$ rather than the spectral shape and the best fits remain
mutually consistent and stable even as the response kernel becomes more sharply peaked
at low $v_\min$ as can be seen in Fig.~\ref{curlyR}.

The SHM reconstruction analysis thus establishes two baseline properties. First, the reconstruction recovers
the slowly varying portion of the input across the accessible $v_\min$ range at both considered DM masses. Second, TES and
MKID detector configurations already display complementarity through their offset but overlapping windows,
and hence that combined measurements between experiments can both broaden the reconstructed halo range and provide  
consistency checks in the overlapping range. By construction the SHM has no significant velocity-dependent
structure in this window and the discriminating power of the method could highlight possible components additional to the SHM.

In Fig.~\ref{DarkDisk} we display the SHM+DD reconstructions for the two masses and two contributing
DM disk density fractions. In each panel the solid curve is the full SHM+DD prediction of
Eq.~\eqref{etaSHMpDD}, the dashed curve depicts pure SHM  and the dash-dotted curve depicts the
SHM scaled by $(1-f_\tx{DD})$ corresponding to the halo function at fixed total density with
the disk component removed. The gap between the dash-dotted and solid curves at low $v_\min$
is the isolated disk contribution $f_\tx{DD} \wt\eta_\tx{DD}$, hence the DD signature
is directly readable from the bin distribution.

For $f_\tx{DD} = 5\%$ case the departure from the pure SHM is found to be small across the
accessible window range, and the best fit bins follow the SHM+DD curve while remaining
close to the SHM behavior. Thus, discriminating quantitatively between the two possibilities would require analysis of the   uncertainty
bands that is beyond our scope. For the more extreme case of $f_\tx{DD} = 25\%$ the excess is clearly more pronounced. At
$m_\chi = 10~\MeV$ the accessible window lies near the peak of $\wt\eta_\tx{DD}$,
and the low $v_\min$ best-fit TES bins rise above the SHM curve, accommodating DD contributions, onto the SHM+DD
curve. For $m_\chi = 1~\GeV$ the window shifts further into the range dominated by disk contributions, at lower $v_\min$. The MKID configuration, with its higher energy threshold and
different TiN response, covers a higher $v_\min$ window and anchors the transition
where SHM and SHM+DD curves converge. Together, the detector configurations constrain both the amplitude of the
excess beyond SHM and the speed range where it drops, determined by $v_0^\tx{DD}$ and $v_e^\tx{DD}$.
The framework therefore could enable, subject to detailed analysis of uncertainties, potentially resolving low $v_\min$ features once their amplitudes are
comparable to or larger than the SHM plateau. The detector complementarity is shown to play a prominent role
especially when the features in halo function are less pronounced or the accessible window ranges are narrow.

In Fig.~\ref{Bound} we display the SHM+EB reconstruction at $m_\chi = 1~\GeV$ on a
logarithmic scale for $\tilde{\eta}$. We restrict to $1~\GeV$ since
$v_{\ast}$ must fall below
$v_\tx{esc}^\tx{EB} \simeq 11.2$~km~s$^{-1}$  for the EB component to noticeably contribute, and at
$10~\MeV$ this lies below the detector thresholds  leaving no accessible bins sensitive to EB components. Reaching lighter masses requires detector configurations with lower thresholds. The model curve rises by about a decade in $\wt\eta$ per
decade in $v_\min$ as $v_\min \to 0$, reflecting the dramatic potential EB overdensity beyond SHM in the region below Earth's
escape speed. Both detector configurations we consider are capable of 
recovering this steep rise within their window ranges without any assumed functional form of the halo model. This highlights
 regimes where the HI approach yields information unavailable to
conventional direct DM detection parametric halo fits. The right panel shows the corresponding per bin event
numbers that peak in the lowest accessible $E'$ bins, the observational
counterpart of the low $v_\min$ enhancement mapped through $\mc R_{ij}$. We stress that the very
large counts follow directly from the assumed
$\rho_\tx{EB} \simeq 10^{14}$~GeV~cm$^{-3}$ overdensity combined with $\sigma_e = 10^{-30}~\cm^2$, which is not
realistic and depicted for illustration.

Across the three considered halo function benchmarks that reflect scenarios respectively with a flat plateau, a modest low $v_\min$ excess and a
steep decade per decade rise, the HI reconstruction is seen to appropriately track the input $\wt\eta$ functions and shows consistent complementarity of detector configurations. In each case the analysis output is a 
best-fit  piecewise constant $\wt\eta$ function,
obtained without assuming a specific halo
ansatz, which carries two key physical information components, related to halo distribution normalization and spectral
shape. The normalization is associated with prefactor $\rho_\chi\sigma_e/m_\chi$ of
Eq.~\eqref{halofuncDef}. For fixed $m_\chi$ and $\rho_\chi$, 
the
plateau level of the best fit $\wt\eta$ function constrains $\sigma_e$ and hence the coupling product
$g_e g_\chi$ that enters the DM-electron amplitude, even if the velocity distribution
departs from the SHM. The spectral shape across bins indicates the underlying model scenario that is nearly flat for the
SHM as shown in Fig.~\ref{SHM}, has a modest low $v_\min$ excess for a DD
 as shown in Fig.~\ref{DarkDisk} and has a steep rise toward $v_\min \to 0$ for an
Earth-bound population  as shown in  Fig.~\ref{Bound}. The best fits already clearly illustrate
different low $v_\min$ trends across the three scenarios, however quantitative model
discrimination requires detailed analysis of uncertainty bands on the $\wt\eta_j$ that we defer to
future work.

The HI reconstruction also carries an incorporated restriction for physical results. Any $f_\chi(v)$
yields a non-increasing $\wt\eta(v_\min)$. Hence, the monotonicity constraint described above is automatically satisfied by a genuine DM signal.
This restriction  requires no additional astrophysical input
and is present in conventional direct DM detection parametric halo analysis fits. Moreover, because
$\wt\eta(v_\min)$ is detector independent  a reconstruction from one experiment can
be propagated to predict spectrum in another experiment. Inserting results from TES-derived $\wt\eta$ analysis into
Eq.~\eqref{mudef} with the MKID response matrix enables predicting the MKID experiment counts, and vice
versa. Thus, the HI method enables a direct cross experiment consistency test as the halo function reconstructed from one experiment can be used to predict the spectrum in another, and agreement with the observed spectrum can be assessed without assuming a specific halo model.

We further comment on the origin of the detector complementarity. The Al target used for the TES benchmark is modeled with a data driven Mermin energy loss function that includes material specific interband and collective structure. By contrast  the TiN target used for the MKID benchmark is described by a smoother finite width Lindhard response containing only the intraband free electron contribution. These differences lead to distinct response functions and sensitivity to different $v_{\min}$ ranges.  Neither the Al nor the TiN plasmon enters the
sub-eV to eV window analyzed here as the Al plasmon resides near $\sim 15~\eV$ and the
free-electron TiN plasmon is several eV above the window. Hence, in both cases the
response is governed by the low energy electron-hole continuum. Through
$v_\min$, the richer momentum structure of the Al energy loss function produces the
multiply peaked TES response of Fig.~\ref{fig:differentialcurlyR}. For TiN the response is comparatively smoother. Hence, the two
detector configurations probe structurally different parts of $\wt\eta(v_\min)$. 

Since binned data may admit degenerate best fit halo functions, a statistically robust cross-experiment prediction requires propagating the corresponding degeneracy or confidence set, or performing a joint fit. In the present analysis we illustrate the mapping using only the 
best-fit solutions.

\subsection{Other Detector Applications} 

The HI formalism described here applies to any sub-eV sensor once its response function
is computed. Notably, this is not restricted to energy resolving
detectors. A threshold (i.e. counting) detector that does not resolve the
deposited energy spectrum still contributes a well defined response function that is determined  by its threshold and target material. If different thresholds can be achieved they could be combined into binned
data. 

One such example is superconducting nanowire single photon detectors (SNSPDs), reaching sub-eV thresholds with low dark count rates while
their superconducting films (e.g. based on NbN or WSi) provide additional target materials. The QROCODILE concept, based on SNSPD readout, reaches very low thresholds and can extend $v_\min$ coverage~\cite{QROCODILE:2024nqm}. Energy resolving platforms such
as SQUID-read magnetic microcalorimeters and bolometers provide a complementary
spectrometric handle. Hence, a multi-target, multi-threshold program combining these
channels is a natural extension of the low $v_\min$ HI approach.

\section{Conclusions}\label{sec:conclusion}

We have presented a HI framework for interpreting sub-GeV DM signals in quantum sensors with sub-eV thresholds. By separating the common halo function $\widetilde{\eta}(v_{\min})$ from the particle physics, material  and detector response, the method enables translating measured experimental spectra into information about the local DM velocity distribution without assuming a specific halo model. 

Using TES (Al) and MKID (TiN) benchmarks, we have shown that different sensor materials and thresholds probe complementary, partially overlapping regions of $v_{\min}$. Their combination can therefore extend the velocity range accessible to reconstruction and provide a direct consistency test. A genuine DM signal must yield a common halo function across detectors. The benchmark SHM, dark disk  and Earth-bound scenarios further illustrate how non-standard low velocity populations could leave distinguishable features in the reconstructed distribution. The detector response functions used here incorporate realistic material and detector inputs. The reconstructions themselves are intended as demonstrations of the method and employ deliberately high signal statistics. Establishing realistic discovery and discrimination sensitivity reach will require detailed confidence bands and a complete treatment of statistical and systematic uncertainties, which we leave for future work.

More broadly, this framework turns low threshold quantum sensors from instruments that only constrain interaction strengths into potential probes of the local DM environment. It can be extended to other energy resolving or threshold counting platforms once their response functions are known. This motivates a multi-material and multi-platform program capable of mapping DM velocities in a regime that is difficult to access with conventional direct DM detection experiments.

%%%%%%%%%%%%%%%%%%%%%%%%%%%%%%%%%%%%%%
\section*{Acknowledgments}
G.B.G.\ is partially supported by the US Department of Energy under Award Number DE-SC0009937. V.T.\ and M.C.\ were supported by the World Premier International Research Center Initiative (WPI), MEXT, Japan.
V.T.\ acknowledges support from JSPS KAKENHI grant No.~23K13109.

%%%%%%%%%%%%%%%%%%%%%%%%%%%%%%%%%%%%%%
\appendix

\section{General Scattering Formalism}\label{changeofvar}

Consider a non-relativistic DM particle of mass $m_\chi$ with incoming momentum $\bs p=m_\chi\bs v$ and outgoing momentum $\bs p^\pr=\bs p-\bs q$, where $\bs q$ is the momentum transferred to the target.

Energy conservation implies that the deposited energy $E_e$ is
\al{
E_e
=E_i-E_f
=\frac{\bs p^2}{2m_\chi}
-\frac{(\bs p-\bs q)^2}{2m_\chi}
=\frac{2\bs p\cdot\bs q-q^2}{2m_\chi},
}
where $E_i=\bs p^2/(2m_\chi)$ and $E_f=\bs p^{\pr 2}/(2m_\chi)$ are the initial and final DM kinetic energies. Thus, 
\al{
E_e
&=\bs q\cdot\bs v-\frac{q^2}{2m_\chi}
=qv\cos\theta-\frac{q^2}{2m_\chi},
\label{Ee_def}
}
where $\theta$ is the angle between $\bs q$ and $\bs v$. 
Solving for the incident speed gives
\al{
v
=\frac{1}{\cos\theta}
\left(
\frac{E_e}{q}+\frac{q}{2m_\chi}
\right) = {v_\min\ov \cos\theta}.
}
For fixed $q$ and $E_e$, the smallest kinematically allowed speed is obtained for $\cos\theta=1$. 

For fixed $v_\min$ and $E_e$, the equation
\al{
E_e=qv_\min-\frac{q^2}{2m_\chi}
}
has the two solutions
\al{
q_-(v_\min,E_e)
&=m_\chi v_\min
-\sqrt{m_\chi^2v_\min^2-2m_\chi E_e},
\nn\ 
q_+(v_\min,E_e)
&=m_\chi v_\min
+\sqrt{m_\chi^2v_\min^2-2m_\chi E_e}.
\label{q+q-}
}
The function $v_\min(q, E_e)$ has a unique minimum $v_*=v_\min(q_*,E_e) =\sqrt{{2E_e}/{m_\chi}}$, at $q_* = \sqrt{2m_\chi E_e}$. Real solutions exist when $v_{\rm min} \geq v_{\ast}$.

Consequently, for any $v_\min$ above the minimum value  there are two momenta $q$, namely $ q_-$ and $q_+$, satisfying $q_-\fn{v_\min} < q_* < q_+\fn{v_\min}$.
Thus, to change the integration variable from $q$ to $v_\min$, an integral over a range $\sqbr{q_i,q_f}$ containing $q_\ast$ must be divided into two branch intervals on which the mapping is one-to-one. Their derivatives are
\al{
\frac{\p q_-}{\p v_\min}
&=
m_\chi-
\frac{m_\chi^2v_\min}
{\sqrt{m_\chi^2v_\min^2-2m_\chi E_e}}
<0,
\nn\
\frac{\p q_+}{\p v_\min}
&=
m_\chi+
\frac{m_\chi^2v_\min}
{\sqrt{m_\chi^2v_\min^2-2m_\chi E_e}} >
0.
}

Defining $v_i = v_\min\fn{q_i}$ and $v_f = v_\min\fn{q_f}$
and reversing the limits on the lower momentum branch gives
\al{
    I &= \int_{q_i}^{q_f}\df q F\fn{q}
    = \int_{q_i}^{q_*}\df q F\fn{q} + \int_{q_*}^{q_f}\df q F\fn{q}\nn
    &= \int_{v_*}^{v_i}\df v_\min {\ab{\p q_-\ov\p v_\min} F\fn{q_-\fn{v_\min}}}\nn
    &\quad+ \int_{v_*}^{v_f}\df v_\min {\ab{\p q_+\ov\p v_\min} F\fn{q_+\fn{v_\min}}}~. 
}
Thus, the absolute Jacobians account for the opposite orientations of the two branches.
 
For the scattering rate integrals considered in this work  the original momentum range is $q\in\sqbr{0,\infty}$, for fixed $E_e>0$  both transformed upper limits are formally infinite. Because the halo function $\wt{\eta}(v_\min)$ vanishes above the maximum speed $v_{\tx{max}}$ supported by the detector frame DM distribution both integrals can instead be truncated at $v_{\tx{max}}$. For the SHM with fixed Earth speed $v_e$, $v_i =v_f=v_{\tx{max}}=v_{\tx{esc}}+v_e$ (which defines $q_i = q_-(v_{\tx{max}})$ and  $q_f = q_+(v_{\tx{max}})$).

\section{Derivation of Event Rate}\label{detailedEventRate} 
 
This derivation follows closely Refs.~\cite{Chen:2021qao,Chen:2022xzi}. Starting from the DM interaction rate defined in Eqs.~\eqref{R-definition}-\eqref{DM form factor},  
imposing energy conservation for the deposited energy $E$, we obtain
\begin{widetext}
\al{
    R &= {\pi \ov \rho_T\mu_{\chi e}^2}{\rho_\chi\ol\sigma_e\ov m_\chi}
    \int\df^3 \bs v f_\chi\fn{\bs v}
    \sqbr{
        \int{\df^3\bs q\ov \pn{2\pi}^3}\ab{\mc F_\med\fn{q}}^2 
        \pn{
            \int_0^\infty \df E~ S\fn{q, E}\delta\fn{E - E_e}
        }
    }
}~.
Changing the order of integration, we can write
\al{
    R &= {\pi\ov \rho_T\mu_{\chi e}^2}
    \int_{0}^{\infty} \df E 
    \int{\df^3\bs q\ov \pn{2\pi}^3}\ab{\mc F_\med\fn{q}}^2 S\fn{q, E}
    \sqbr{
        {\rho_\chi\ol\sigma_e\ov m_\chi}
        {
            \int\df^3 \bs v f_\chi\fn{\bs v}
            \delta\fn{E - E_e}
        }
    }
    = \int_{0}^{\infty} \df E {\df R\ov \df E}\fn{E},
}
that defines the differential event rate ${\df R/ \df E}$. 

As we have discussed in App.~\ref{changeofvar}, when we fix the magnitude of momentum transfer $q$, there is a unique velocity minimum $v_\min$ that realizes the particular energy deposit $E$. Namely,  
\al{
    \delta\fn{E - E_e}
    = \delta\fn{E - \sqbr{qv\cos\theta - {q^2\ov 2 m_\chi}}}
    = {1\ov qv}\delta\fn{\cos\theta - {v_\min\ov v}}.
}
Using this delta function to perform the angular integration in $\df^3\bs q$, we obtain
\al{
    {\df R\ov \df E}\fn{E}
    &= 
    {1\ov {4\pi}\rho_T\mu_{\chi e}^2}
    \int_{0}^{\infty}\df q~
    q\ab{\mc F_\med\fn{q}}^2 S\fn{q, E}\sqbr{
        {\rho_\chi\ol\sigma_e\ov m_\chi}
        {
            \int\df^3 \bs v 
            {f_\chi\fn{\bs v}\ov v}\Theta\fn{v - v_\min}
        }
    }.
}
With the halo function $\wt\eta\fn{v_\min\fn{q, E}}$, 
\al{
    {\df R\ov \df E}\fn{E}
    &= 
    {1\ov {4\pi}\rho_T\mu_{\chi e}^2}
    \int_{0}^{\infty} \df q~
    q\ab{\mc F_\med\fn{q}}^2 S\fn{q, E}
    \wt\eta\fn{v_\min\fn{q, E}}.
}
\end{widetext}
Using the two momentum transfer branches derived in App.~\ref{changeofvar}, the $q$ integral can then be transformed into the $v_\min$ integral given in Eq.~\eqref{pureRate}.

\section{Truncated Maxwellian Halo Function}\label{Local-DM}

Each benchmark model of Sec.~\ref{benchmarks} makes use of the truncated Maxwellian velocity
distribution of the form of Eq.~\eqref{SHM_dif}, differing only in the parameters
$(v_0, v_e, v_\tx{esc})$ and the density prefactor. Here, we state the
corresponding closed form halo function expression, obtained by substituting
Eq.~\eqref{SHM_dif} into the definition Eq.~\eqref{halofuncDef} and carrying out
the angular integration.

For parameters $(v_0, v_e, v_\tx{esc})$ and density $\rho$,
\begin{widetext}
\al{
\wt\eta(v_\min)
&= \frac{\rho~\sigma_e}{m_\chi}~
\frac{\pi v_0^2}{2 v_e~ K(v_0, v_\tx{esc})}\nn
&\quad\times
\begin{cases}
\displaystyle
-4 e^{-v_\tx{esc}^2/v_0^2}~ v_e
+\sqrt{\pi}~ v_0
\left[
\erf\!\left(\frac{v_\min+v_e}{v_0}\right)
-\erf\!\left(\frac{v_\min-v_e}{v_0}\right)
\right],
& v_\min < v_\tx{esc} - v_e, \\[2ex]
\displaystyle
-2 e^{-v_\tx{esc}^2/v_0^2}~(v_e+v_\tx{esc}-v_\min)
+\sqrt{\pi}~ v_0
\left[
\erf\!\left(\frac{v_\tx{esc}}{v_0}\right)
-\erf\!\left(\frac{v_\min-v_e}{v_0}\right)
\right],
& v_\tx{esc} - v_e < v_\min < v_\tx{esc} + v_e, \\[2ex]
0, & v_\min > v_\tx{esc} + v_e,
\end{cases}
\label{eta_closed}
}
\end{widetext}
where the truncated-Maxwellian normalization factor is
\al{
K(v_0, v_\tx{esc}) =
\pi^{3/2} v_0^3
\left[
\erf\!\left(\frac{v_\tx{esc}}{v_0}\right)
-\frac{2}{\sqrt{\pi}}~\frac{v_\tx{esc}}{v_0}~
e^{-v_\tx{esc}^2/v_0^2}
\right].
\label{Kfactor}
}
The benchmark halo model functions then follow directly.  For the Earth-bound benchmark  for which $v_e^{\rm EB}=0$  Eq.~\eqref{eta_closed} is understood in the smooth $v_e\to0$ limit.

\clearpage

\bibliography{biblio.bib}

\end{document}